\documentclass[prd,aps,showpacs,floats,letterpaper,floatfix,
groupedaddress,superscriptaddress
,nofootinbib
,eqsecnum
,twocolumn
]{revtex4}
\usepackage{amsmath}
\usepackage{amsfonts}
\usepackage{bm}
\usepackage{graphicx}

\def\v1v2{{\bf v}_1 \cdot {\bf v}_2}
\allowdisplaybreaks

\begin{document}

\title{Orbital flips in hierarchical triple systems: relativistic effects and third-body effects to hexadecapole order}

\author{
Clifford M.~Will} \email{cmw@phys.ufl.edu}
\affiliation{Department of Physics, University of Florida, Gainesville, Florida 32611, USA}
\affiliation{GReCO, Institut d'Astrophysique de Paris, CNRS,\\ 
Universit\'e Pierre et Marie Curie, 98 bis Boulevard Arago, 75014 Paris, France}

\date{\today}

\begin{abstract}
We analyze the secular evolution of hierarchical triple systems in the post-Newtonian approximation to general relativity.    We expand the Newtonian three-body equations of motion in powers of the ratio $a/A$, where $a$ and $A$ are  the semimajor axis of the inner binary's orbit  and of the orbit of the third body relative to the center of mass of the inner binary, respectively.    The leading order ``quadrupole'' terms, of order $(a/A)^3$ relative to the $1/a^2$ acceleration within the inner binary, are responsible for the well-known Kozai-Lidov oscillations of orbital inclination and eccentricity.  The octupole terms, of order $(a/A)^4$ have been shown to allow the inner orbit to ``flip'' from prograde relative to the outer orbit to retrograde and back, and to permit excursions to very large eccentricities.   
We carry the expansion of the equations of motion to hexadecapole order, corresponding to contributions of order $(a/A)^5$.   We also include the leading orbital effects of post-Newtonian theory, namely the pericenter precessions of the inner and outer orbits.   Using the Lagrange planetary equations for the orbit elements of both binaries, we average over orbital timescales, obtain the equations for the secular evolution of the elements through hexadecapole order, and employ them to analyze cases of astrophysical interest.   We find that, for the most part, the orbital flips found at octupole order are robust against both relativistic and hexadecapole perturbations.    We show that, for equal-mass inner binaries, where the octupole terms vanish, the hexadecapole contributions can alone generate orbital flips and excursions to very large eccentricities.
   
\end{abstract}

\maketitle

\section{Introduction and summary}
\label{sec:intro}

The hierarchical three-body problem, in which a close binary system is in orbit with a distant third body, is as old as Newton's gravity, but continues to yield surprises.   The first surprise came when Newton himself failed to account for the advance of the lunar perigee caused by the perturbing effect of the distant Sun (although a correct calculation does exist in his unpublished papers).   Clairaut published a correct solution in 1749.  Another notable surprise was LeVerrier's failure in 1859 to account for the advance of Mercury's perihelion by including perturbations of the Sun-Mercury binary system due to the distant planets.  The solution to this surprise was famously provided by Einstein.  

A contemporary surprise was the discovery in the 1960s of the Kozai-Lidov mechanism, 
in which, over long timescales, there is an interchange between the eccentricity of the two-body inner orbit and its inclination relative to the plane of the third body.  This remarkable effect was discovered independently by Lidov \cite{1962P&SS....9..719L}, who was studying orbits of artifical satellites, and Kozai \cite{1962AJ.....67..591K} who was studying asteroid orbits.   For many years, interest in the Kozai-Lidov effect was largely confined to solar-system research, until the discoveries of unusual exoplanet and multiple star systems brought the phenomenon into the astrophysical realm.  Because the mechanism could generate orbits with high eccentricity, it even became of interest for general relativistic astrophysics because of the possible enhancement of relativistic effects such as the pericenter advance and the emission of gravitational radiation.    

The Kozai-Lidov mechanism is obtained by expanding the perturbing acceleration in the inner binary's motion caused by the third body in powers of $\epsilon \equiv a/A$, the ratio of the two semimajor axes, and keeping the leading term, which is proportional to $\epsilon^3$ (conventionally called the ``quadrupole order'' term).  A similar expansion is performed on the acceleration of the third body.  The equations of motion are averaged over time to suppress periodic effects and to reveal the long-timescale, secular changes in the orbits.  One immediate consequence is that the two semimajor axes, $a$ and $A$ do not experience secular changes.   In the limit where one of the inner bodies is much less massive than the other, the component of the angular momentum of the inner orbit that is perpendicular to the plane of the outer orbit turns out to be constant.   Since this component is proportional to $\cos z \sqrt{1-e^2}$, where $z$ is the inclination angle between the normals to the two orbital planes and $e$ is the eccentricity of the inner orbit, we see that, as $z$ decreases, $e$ increases, and vice versa.   The variables $e$ and $z$ oscillate between well-defined maxima and minima, depending on the initial conditions.   In addition, if $z$ is initially less than $90^{\rm o}$, so that the inner orbit is prograde relative to the outer orbit, the orbit stays prograde.  If the inner orbit is initially retrograde ($z >  90^{\rm o}$), it stays retrograde.   The inner orbit cannot ``flip'' relative to the outer orbit.

The next surprise came in 2011.   It was known by then that in about 25 percent of exoplanet systems with ``hot Jupiters'', that is Jovian-mass planets close to the host star, the planet was in a retrograde orbit relative to the spin of the star.   If, as in the solar system, the star and the other planets rotated in the same direction, how did these Jupiters end up in retrograde orbits?   Naoz {\em et al.}\ \cite{2011Natur.473..187N} pointed out that, if one included the terms in the perturbing acceleration at the {\em next} order in $\epsilon$, namely $\epsilon^4$,  (called ``octupole order'' terms) then orbital flips could occur.   In addition, unlike the modest variations in eccentricity allowed by the Kozai-Lidov mechanism, excursions to eccentricities very close to unity could occur.  (These behaviors had actually been noticed almost a decade earlier 
\cite{1999MNRAS.304..720K,2000ApJ...535..385F,2002ApJ...578..775B}, but at the time there was no obvious astrophysical application.)   As a result a ``run of the mill'' Jupiter, perturbed by a more distant planet, could be flipped to a retrograde orbit and also brought very close to the star, where tidal and other dissipative processes could circularize the orbit, thus producing a retrograde ``hot Jupiter''.

In follow-up papers, Naoz and collaborators \cite{2013MNRAS.431.2155N,2014ApJ...785..116L} studied other situations in which orbital flips could occur.  Naoz {\em et al.}\ \cite{2013ApJ...773..187N} studied the effects of post-Newtonian general relativistic (GR) corrections, including gravitational radiation reaction, on the generation of orbital flips and extreme eccentricities, while Liu {\em et al.}\ \cite{2015MNRAS.447..747L} studied the impact of short-range forces induced by tidal, rotational and GR effects on these extreme phenomena.   Lithwick and Naoz \cite{2011ApJ...742...94L} and Katz {\em et al.} \cite{2011PhRvL.107r1101K} studied the case where one of the inner bodies is a ``test'' particle.  Naoz \cite{2016ARA&A..54..441N} provides a thorough review of these effects in hierarchical triple systems and discusses their astrophysical implications.  

Given the complexity of the hierarchical three-body problem, it is natural to ask, are there more surprises?   To that end, we have gone to the next order in the expansion of the perturbing acceleration, to order $\epsilon^5$, called ``hexadecapole order''.    Other authors have addressed this level of approximation in a range of contexts, mainly using the canonical approach of Delaunay variables.
 Laskar and Bou\'e  \cite{2010A&A...522A..60L} obtained the disturbing function in the Hamiltonian formally to all orders and explicitly to very high orders in $\epsilon$; they did not derive the explicit equations of motion at hexadecapole order.   Hamers derived the secular equations through hexadecapole order (unpublished), and Hamers and Portegies Zwart 
\cite{2016MNRAS.459.2827H} expanded the Hamiltonian for an  $N$-body system in a sequence of hierarchical orbits to hexadecapole and dotriocontupole ($\epsilon^6$) orders.  Antognini
 \cite{2015MNRAS.452.3610A} derived (though did not display) the secular equations through hexadecapole order in both the Delaunay approach and in a method using eccentricity and angular momentum vectors, and made the code publicly available.   Carvalho {\em et al.}\ 
\cite{2016CeMDA.124...73C} derived the contributions to the disturbing function at hexadecapole and dotriocontupole orders, but under the assumption that the orbital plane of the third body is fixed. 

We use the approach of ``osculating orbit elements'' whereby each of the orbits is characterized by its semimajor axis  and eccentricity, its inclination and angle of ascending node relative to a reference coordinate system, and its angle of pericenter measured from the ascending node.  The equations of motion for the two orbits can then be rewritten as the ``Lagrange planetary'' equations for the orbit elements, which take the generic form $dY^\alpha/dt = Q^\alpha (X^\beta, Z^\gamma, t)$, where  
$X^\beta$ and $Z^\gamma$ denote orbit elements of the inner and outer binary, respectively, and $Y^\alpha$ denotes an orbit element of either binary.   We then carry out the conventional average over an orbit of both the inner binary and the outer binary, arriving at equations for the secular changes in the orbit elements.   
To quadrupole and octupole orders, our equations for the secular evolution of the elements agree completely with those derived using the Delaunay variables approach, and presented in Sec.\ 2.2 of Ford {\em et al.}\ \cite{2000ApJ...535..385F}, or in Appendix A and B of  Naoz {\em et al.}\ \cite{2013MNRAS.431.2155N}.   

We incorporate the effects of general relativity (GR) by adding to the secular equations the relativistic pericenter advances of both the inner and outer orbits at the first post-Newtonian order (we do not include additional GR terms studied in \cite{2013ApJ...773..187N}).   
We then apply these hexadecapole-order equations including GR to a number of case studies presented in the literature, particularly those where orbital flips and large eccentricity excursions occur at octupole order \cite{2011ApJ...742...94L,2013MNRAS.431.2155N,2014ApJ...785..116L}.    We also explore the special case where the masses comprising the inner binary are equal.  In this case the octupole terms vanish identically.  Nevertheless, we find a ``sweet spot'' in the space of orbits where the hexadecapole terms alone can generate orbital flips and large eccentricity excursions. 

The remainder of this paper presents details.  In Sec.\ \ref{sec:secular} we present the detailed derivation of the secular Lagrange planetary equations through hexadecapole order.  In Sec.\ \ref{sec:numerical} we present five case studies analyzed using these higher-order equations. Section \ref{sec:equalmass} considers the equal-mass case.    Section \ref{sec:conclusions} presents concluding remarks.  In an Appendix, we present a dictionary for converting between the language of osculating orbit elements and the Delaunay variables approach used in  \cite{2000ApJ...535..385F,2013MNRAS.431.2155N}.   Hereafter, we refer to the two papers by Naoz {\em et al.}\ 
\cite{2011Natur.473..187N,2013MNRAS.431.2155N} collectively as NFLRT.

\section{Secular evolution of hierarchical triple systems}
\label{sec:secular}

\subsection{Equations of motion and conserved quantities}

We begin with the Newtonian equations of motion for a three-body system,
\begin{equation}
\bm{a}_a = - \frac{Gm_b \bm{x}_{ab}}{r_{ab}^3}  - \frac{Gm_c \bm{x}_{ac}}{r_{ac}^3} \,,
\end{equation}
where $a = (1,\,2,\,3)$, $b \ne c \ne a$, $G$ is Newton's constant, $\bm{x}_{ab} \equiv \bm{x}_a - \bm{x}_b$, and $r_{ab} \equiv |\bm{x}_{ab}|$.   Bodies 1 and 2 will be taken to be the ``inner'' binary, with body 3 taken to be the ``outer'' perturbing body.   We define the centers of mass of the system and of the inner binary to be
\begin{align}
\bm{X}_c &\equiv \frac{1}{M} \left (m_1 \bm{x}_1 +m_2 \bm{x}_2+m_3 \bm{x}_3 \right ) \,,
\nonumber \\
\bm{x}_c &\equiv  \frac{1}{m} \left (m_1 \bm{x}_1 +m_2 \bm{x}_2 \right ) \,,
\label{eq2:eom}
\end{align}
where $M \equiv m_1  +m_2 +m_3$ and $m \equiv m_1  +m_2$.
A ``hierarchical'' triple system is one in which the orbital separation of the inner binary is small compared to that of the outer binary, so we expand the equations of motion in powers of that small ratio.  This can be carried out by writing $\bm{x}_{a3} =  \bm{x}_{ac} + \bm{x}_{c3}$, where $a = (1, \, 2)$, with $|\bm{x}_{ac}| \ll |\bm{x}_{c3}|$, and using the Taylor expansion
\begin{equation}
\frac{{x}_{a3}^j}{r_{a3}^3} = \frac{{x}_{c3}^j}{r_{c3}^3} - \sum_{\ell =1}^\infty \frac{1}{\ell !} x_{ac}^L \partial_c^{jL} \left ( \frac{1}{r_{c3}}  \right ) \,,
\label{eq2:taylor}
\end{equation}
where the superscript $L$ is a multi-index, with the interpretation $z^L \equiv z^{i_1} z^{i_2} \dots z^{i_\ell}$; similarly, $\partial_c^{jL}$ is a multi-partial derivative with respect to $\bm{x}_c$, and a contraction over the $\ell$ repeated indices is assumed.   We now define $\bm{x} \equiv \bm{x}_1 - \bm{x}_2$, $r \equiv |\bm{x}|$, $\bm{n} \equiv \bm{x}/r$, $\bm{X} \equiv \bm{x}_{3c}$, $R \equiv |\bm{X}|$, $\bm{N} \equiv \bm{X}/R$; note that $\bm{X}$ is chosen to point from the inner binary to the third body.  We also define the dimensionless mass coefficients $\alpha_i \equiv m_i/m$, ($i = 1,\,2$), with $\alpha_1 +\alpha_2 =1$.
We define the dimensionless mass difference and dimensionless reduced mass 
\begin{align}
\Delta &\equiv \alpha_2 - \alpha_1 \,,
\nonumber \\
\eta &\equiv \alpha_1 \alpha_2 \,,
\label{eq2:Deltaeta}
\end{align}
and choose body 1 to be less than or equal in mass to body 2, so that $\Delta = \sqrt{1-4\eta} \ge 0$; recall that $\eta$ ranges between zero and $1/4$.   We note that $\bm{x}_{1c} = \alpha_2 \bm{x}$ and $\bm{x}_{2c} = - \alpha_1 \bm{x}$.  With this convention, in the limit $m_1 \to 0$, $\eta \to 0$, $\Delta \to 1$, and the relative orbit $\bm x$ and the actual orbit ${\bm x}_{1c}$ of body 1 coincide.  We also recall that $\partial_c^{jL} r_{c3} = (2\ell +1)!! N^{\langle jL \rangle}/R^{\ell +2}$, where the superscript $\langle \dots \rangle$ denotes a symmetric tracefree product (for a review see \cite{PW2014}).   

We can then express the equation of motion of the inner binary and of the third body relative to the inner center of mass in the general form
\begin{align}
a^j &= - \frac{Gm n^j}{r^2} + \frac{Gm_3 }{R^2} \sum_{\ell = 1}^\infty \frac{(2\ell +1)!!}{\ell !} \left ( \frac{r}{R} \right )^\ell 
\nonumber \\
& \qquad \times 
n^L N^{\langle jL \rangle} \left [ \alpha_2^\ell - (-\alpha_1)^\ell \right ] \,,
\nonumber \\
A^j &=-  \frac{GM N^j}{R^2} -  \eta\frac{GMr}{R^3} \sum_{\ell = 1}^\infty \frac{(2\ell +3)!!}{(\ell+1) !} \left ( \frac{r}{R} \right )^\ell 
\nonumber \\
& \qquad \times
n^{L+1} N^{\langle j(L+1) \rangle} \left [ \alpha_2^{\ell} - (-\alpha_1)^{\ell} \right ] \,,
\label{eq2:eom2}
\end{align}
where $\bm{a} \equiv d^2 \bm{x}/dt^2$ and $\bm{A} \equiv d^2 \bm{X}/dt^2$.  Note that the perturbing terms in the equation for $A^j$ depend on the inner binary's reduced mass parameter $\eta$; this is to be expected, since in the limit in which body 1 is a test body, $\eta \to 0$, and the third body moves on an unperturbed Keplerian orbit around the massive body 2.   

The equations admit conserved total energy and angular momentum, given by
\begin{align}
{\cal E} &= \frac{1}{2} \sum_a m_a v_a^2 - \frac{1}{2} \sum_{a,b} \frac{Gm_a m_b}{r_{ab}} 
\nonumber \\
&= \frac{1}{2} \left ( m\eta v^2 + M\eta_3 V^2 \right ) - \frac{Gm^2 \eta}{r}  - \frac{GM^2 \eta_3}{R}
\nonumber \\
& \qquad  - \eta \, \eta_3 \frac{GM^2  r}{R^2}  \sum_{\ell = 1}^\infty \frac{(2\ell +1)!!}{(\ell +1) !} \left ( \frac{r}{R} \right )^\ell 
\nonumber \\
& \qquad \times
n^{L+1} N^{\langle L +1 \rangle} \left [ \alpha_2^{\ell} - (-\alpha_1)^{\ell} \right ] \,,
 \nonumber \\
\bm{J} &= \sum_a m_a \bm{x}_a \times \bm{v}_a 
\nonumber \\
& 
= m \eta \left (\bm{x} \times \bm{v} \right )
 + M \eta_3 \left (\bm{X} \times \bm{V} \right ) \,,
 \label{eq2:conserved}
\end{align}
where $\eta_3 \equiv m m_3/M^2$, and we have chosen the coordinates so that $\bm{X}_c = 0$.

Beginning with $\ell =1$, the terms in the
expansions over $\ell$ are conventionally denoted quadrupole, octupole, hexadecapole, dotriocontupole, etc.  We expand the equations through hexadecapole order, leading to the final forms
\begin{align}
a^j &= - \frac{Gm n^j}{r^2} + \frac{Gm_3 r}{R^3} \left ( 3 N^j N_n - n^j \right ) 
\nonumber \\
& \quad
+ \frac{3}{2} \frac{Gm_3 r^2}{R^4} \Delta \left ( 5 N^j N_n^2 - 2 n^j N_n - N^j  \right )  
\nonumber \\
& \quad
 + \frac{1}{2}  \frac{Gm_3 r^3}{R^5} (1-3\eta) \left ( 35 N^j N_n^3 - 15 n^j N_n^2 
\right .
\nonumber \\
& \qquad 
\left .
- 15 N^j N_n +3n^j  \right ) \,,
\nonumber \\
A^j &=-  \frac{GM N^j}{R^2} - \frac{3}{2} \frac{GM r^2}{R^4} \eta \left ( 5 N^j N_n^2 - 2 n^j N_n - N^j  \right )
\nonumber \\
& \quad
- \frac{1}{2}  \frac{GM r^3}{R^5} \eta \Delta \left ( 35 N^j N_n^3 - 15 n^j N_n^2 
\right .
\nonumber \\
& \qquad 
\left .
- 15 N^j N_n +3n^j  \right ) 
\nonumber \\
& \quad
- \frac{5}{8}  \frac{GM r^4}{R^6} \eta  (1-3\eta) \left ( 63 N^j N_n^4 - 28 n^j N_n^3 
\right .
\nonumber \\
& \qquad 
\left .
- 42 N^j N_n^2 +12 n^j N_n +3N ^j  \right )  \,,
\label{eq2:eom3}
\end{align}
where $N_n \equiv \bm{N} \cdot \bm{n}$.   The octupole perturbations depend on the dimensionless mass difference $\Delta$, while the hexadecapole perturbations depend on the factor $(1-3\eta)$.    
To hexadecapole order, the energy is given by
\begin{align}
{\cal E} &=  \frac{1}{2} \left ( m\eta v^2 + M\eta_3 V^2 \right ) - \frac{Gm^2 \eta}{r}  - \frac{GM^2 \eta_3}{R}
\nonumber \\
&\quad
- \frac{1}{2} \frac{GM^2 \eta \eta_3 r^2}{R^3} \biggl [ (3N_n^2 -1) 
+ \Delta \left (\frac{r}{R} \right )  N_n \left ( 5N_n^2 - 3 \right )
\nonumber \\
& \qquad 
+ \frac{1}{4} (1-3\eta) \left (\frac{r}{R} \right )^2  (35 N_n^4 - 30 N_n^2 +3)
\biggr ] \,.
\label{eq2:energy}
\end{align}

\subsection{Osculating orbit elements and the Lagrange planetary equations}

We define the osculating orbit elements of the inner and outer orbits in the standard manner: for the inner orbit, we have the orbit elements $p$, $e$, $\omega$, $\Omega$ and $\iota$, with the definitions
\begin{eqnarray}
r &\equiv& p/(1+e \cos f) \,,
\nonumber \\
{\bm x} &\equiv& r {\bm n} \,,
\nonumber \\
{\bm n} &\equiv& \left [ \cos \Omega \cos(\omega + f) - \cos \iota \sin \Omega \sin (\omega + f) \right ] {\bm e}_X 
\nonumber \\
&&
 + \left [ \sin \Omega \cos (\omega + f) + \cos \iota \cos \Omega \sin(\omega + f) \right ]{\bm e}_Y
\nonumber \\
&&
+ \sin \iota \sin(\omega + f) {\bm e}_Z \,,
\nonumber \\
{\bm \lambda} &\equiv& d{\bm n}/df \,, \quad \hat{\bm h}={\bm n} \times {\bm \lambda} \,,
\nonumber \\
{\bm h} &\equiv& {\bm x} \times {\bm v} \equiv \sqrt{Gmp} \, \bm{\hat{h}} \,,
\label{eq2:keplerorbit1}
\end{eqnarray}
where (${\bm e}_X,\,{\bm e}_Y ,\,{\bm e}_Z$) define a reference basis, to be specified below.  From the given definitions, it is evident that ${\bm v} = \dot{r} {\bm n} + (h/r) {\bm \lambda}$ and $\dot{r} = (he/p) \sin f$. 

For the outer orbit, we have the elements $P$, $E$, $\omega_3$, $\Omega_3$, and $\iota_3$, with the definitions
\begin{eqnarray}
R &\equiv& P/(1+E \cos F) \,,
\nonumber \\
{\bm X} &\equiv& R {\bm N} \,,
\nonumber \\
{\bm N} &\equiv& \left [ \cos \Omega_3 \cos(\omega_3 + F) 
\right .
\nonumber \\
&& \quad
\left .
- \cos \iota_3 \sin \Omega_3 \sin (\omega_3 + F) \right ] {\bm e}_X 
\nonumber \\
&&
 + \left [ \sin \Omega_3 \cos(\omega_3 + F) 
 \right .
\nonumber \\
&& \quad
\left .
+ \cos \iota_3 \cos \Omega_3 \sin (\omega_3 + F) \right ]{\bm e}_Y
\nonumber \\
&&
+ \sin \iota_3 \sin(\omega_3 + F) {\bm e}_Z \,,
\nonumber \\
{\bm \Lambda} &\equiv& d{\bm N}/dF \,, \quad \hat{\bm H}={\bm N} \times {\bm \Lambda} \,,
\nonumber \\
{\bm H} &\equiv& {\bm X} \times {\bm V} \equiv \sqrt{GMP} \, \bm{\hat{h}} \,.
\label{eq2:keplerorbit2}
\end{eqnarray}
In a similar manner, ${\bm V} = \dot{R} {\bm N} + (H/R) {\bm \Lambda}$ and $\dot{R} = (HE/P) \sin F$.
The semimajor axes of the two orbits are defined by $a \equiv p/(1-e^2)$ and $A \equiv P/(1-E^2)$.

 \begin{figure}[t]
\begin{center}

\includegraphics[width=3.4in]{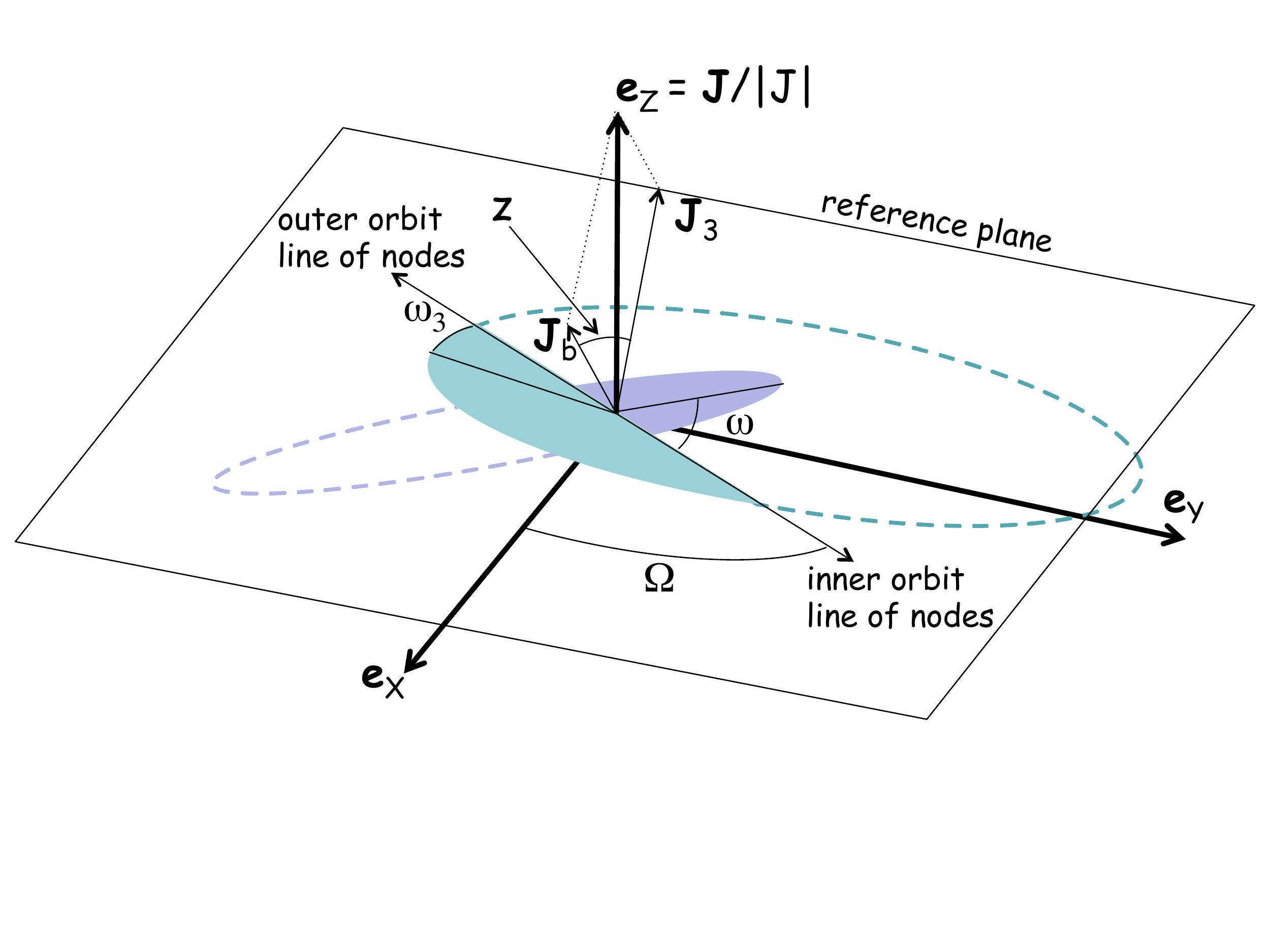}

\caption{Orientation of inner and outer orbits. (Color figures in online version.)
\label{fig:orbits} }
\end{center}
\end{figure}

The total angular momentum is strictly conserved if the system is isolated and we ignore gravitational radiation reaction, therefore it is natural to choose the $Z$-axis to lie along $\bm{J}$, i.e., $\bm{e}_Z = \bm{J}/J$ (see Fig.\ \ref{fig:orbits}).  In general, 
\begin{align}
\bm{J} &= J_b \left [ \sin \iota \left ( \sin \Omega \bm{e}_X - \cos \Omega \bm{e}_Y \right ) + \cos \iota \bm{e}_Z \right ]
\nonumber \\
& \quad + J_3 \left [ \sin \iota_3 \left ( \sin \Omega_3 \bm{e}_X - \cos \Omega_3 \bm{e}_Y \right ) + \cos \iota_3 \bm{e}_Z \right ] \,,
\label{eq2:Jtotal}
\end{align}
where $J_b = m \eta \sqrt{Gmp}$ and $J_3 = M \eta_3 \sqrt{GMP}$.
Thus, to align $\bm{J}$ with the $Z$-axis, we must impose $J_b  \sin \iota  \sin \Omega = - J_3 \sin \iota_3  \sin \Omega_3$ and
$J_b  \sin \iota  \cos \Omega = - J_3 \sin \iota_3  \cos \Omega_3$; this implies that $\tan \Omega = \tan \Omega_3$ and
${\rm sgn} (\sin \Omega) = - {\rm sgn} (\sin \Omega_3)$.  Together, these imply that
\begin{equation}
\Omega_3 = \Omega + \pi \,.
\end{equation}
Another way of stating this result is that the components of the angular momenta of the two orbits in the $X-Y$  plane must be equal and opposite, and thus that the orbital planes must intersect the $X-Y$ plane along a common line, and the lines of ascending nodes must be parallel and in opposite directions.  We then have that $J_b \sin \iota = J_3 \sin \iota_3$.  Defining 
\begin{align}
\beta &\equiv \frac{J_b}{J_3} = \frac{\sin \iota_3}{\sin \iota} \,,
\nonumber \\
z &= \iota + \iota_3 \,,
\label{eq2:betaz}
\end{align}
it is straightforward to obtain the relations
\begin{equation}
\cot \iota = \frac{\beta + \cos z}{\sin z} \,,  \qquad \cot \iota_3 = \frac{\beta^{-1} + \cos z}{\sin z} \,.
\label{eq2:inclinations}
\end{equation}
It will turn out that only the {\em relative} inclination $z$ between the two orbits is dynamically relevant; given an evolution for $z$ and $\beta$, the individual orbital inclinations can be recovered algebraically from Eqs.\ (\ref{eq2:inclinations}).

From Eqs.\ (\ref{eq2:eom3}), we define the perturbing accelerations $\delta {\bm a} \equiv \bm{a} + Gm{\bm n}/r^2$ and $\delta {\bm A} \equiv \bm{A} + GM{\bm N}/R^2$.
Then, for the inner binary, we define the radial $\cal R$, cross-track $\cal S$ and out-of-plane $\cal W$ components of the perturbing acceleration $\delta {\bm a}$, defined respectively by
${\cal R} \equiv {\bm n} \cdot \delta {\bm a}$,
 ${\cal S} \equiv {\bm \lambda} \cdot \delta {\bm a}$ and
 ${\cal W} \equiv \bm{\hat{h}} \cdot \delta {\bm a}$,
and we write down the ``Lagrange planetary equations'' for the evolution of the orbit elements,
\begin{eqnarray}
\frac{dp}{dt} &=& 2 \sqrt{\frac{p^3}{Gm}} \frac{{\cal S}}{1+e \cos f} \,,
\nonumber \\
\frac{de}{dt} &=& \sqrt{\frac{p}{Gm}} \left [ \sin f \, {\cal R} + \frac{2\cos f + e +e\cos^2 f}{1+ e\cos f} {\cal S} \right ]\,,
\nonumber \\
\frac{d\varpi}{dt} &=& \frac{1}{e}\sqrt{\frac{p}{Gm}} \left [ -\cos f \, {\cal R} + \frac{2 + e\cos f}{1+ e\cos f}\sin f {\cal S} 
\right ] \,,
\nonumber \\
\frac{d\iota}{dt} &=& \sqrt{\frac{p}{Gm}} \frac{\cos (\omega +f)}{1+ e\cos f} {\cal W} \,,
\nonumber \\
\sin \iota \frac{d\Omega}{dt} &=& \sqrt{\frac{p}{Gm}} \frac{\sin (\omega +f)}{1+ e\cos f} {\cal W} \,.
\label{eq2:lagrange}
\end{eqnarray}
The auxiliary variable $\varpi$ is defined such that the change in pericenter angle is given by $\dot{\omega} = \dot{\varpi} -  \dot{\Omega} \cos \iota$.

For the outer binary, the analogous components of the perturbing acceleration $\delta {\bm A}$, are defined by
${\cal R}_3 \equiv {\bm N} \cdot \delta {\bm A}$,
 ${\cal S}_3 \equiv {\bm \Lambda} \cdot \delta {\bm A}$ and
 ${\cal W}_3 \equiv \bm{\hat{H}} \cdot \delta {\bm A}$.   
The planetary equations for the outer binary take the form of Eqs.\ (\ref{eq2:lagrange}), with suitable replacements of all the relevant variables, $p \to P$, $e \to E$, $m \to M$, $f \to F$, and so on, and with $\dot{\omega}_3 = \dot{\varpi}_3 - \dot{\Omega}_3 \cos \iota_3 $.

Combining these equations and inserting the perturbing accelerations, it is straightforward to verify directly that
\begin{align}
\frac{d}{dt}  & \left ( \Omega - \Omega_3 \right ) = 0 \,,
\nonumber \\
\frac{d}{dt} &\left ( m\eta \sqrt{Gmp} \sin \iota - M\eta_3 \sqrt{GMP} \sin \iota_3 \right ) = 0 \,,
\nonumber \\
\frac{d}{dt} &\left ( m\eta \sqrt{Gmp} \cos \iota + M\eta_3 \sqrt{GMP} \cos \iota_3 \right ) = 0 \,,
\label{eq2:Jtests}
\end{align}
reflecting the conservation of the three components of the total angular momentum.   

\subsection{Secular evolution of the orbit elements}

We now use first-order perturbation theory to obtain the secular evolutions of the orbital elements.  This is done by substituting constant values of the orbit elements into the right-hand sides of the planetary equations, and averaging over an orbit of both the inner and outer binaries.   This is justified by the fact that the leading order perturbation is at $\epsilon^3$ and we are going out to order $\epsilon^5$.  Were we including terms of dotriocontupole order ($\epsilon^6$) in the equations of motion, for example, it would have been necessary to invoke second-order perturbation theory for the quadrupole terms, in which the full (secular plus periodic) solutions at quadrupole order are substituted back into the Lagrange planetary equations and the orbital average carried out again. 

Each planetary equation can be written in the generic form
\begin{align}
\frac{dY^\alpha}{dt} &= Q^\alpha (X^\beta, Z^\gamma, t) 
= A^\alpha (X^\beta, f) B^\alpha (Z^\gamma, F) \,,
\end{align}
where the $X^\beta$ and $Z^\beta$ are orbit elements associated with the inner and outer binaries, respectively, and where the last step recognizes that every term on the right-hand side can be factorized into a product of terms depending only on one or the other orbital elements and on either $f$ or $F$.  Then the average of $dY^\alpha/dt$ is approximated as product of averages of $A^\alpha$ and $B^\alpha$, in other words
\begin{align}
\left \langle \frac{dY^\alpha}{dt} \right \rangle &= \left \langle A^\alpha \right \rangle \left \langle B^\alpha \right \rangle
\nonumber \\
&
= \frac{1}{P_{\rm inner}} \int_0^{P_{\rm inner}} A^\alpha dt  \,\frac{1}{P_{\rm outer}} \int_0^{P_{\rm outer}} B^\alpha dt \,,
\end{align}
where the two orbital periods are given by $P_{\rm inner} = 2\pi \sqrt{a^3/Gm}$ and $P_{\rm outer} = 2\pi \sqrt{A^3/GM}$, with the assumption that $P_{\rm inner} \ll P_{\rm outer}$.
In integrating over an orbit of the inner binary, it is useful to convert the angular variable from the true anomaly $f$, to the eccentric anomaly $u$, using the relations
\begin{equation}
\sin f = \frac{\sqrt{1-e^2} \sin u}{1- e \cos u} \,, \quad \cos f = \frac{\cos u - e}{1- e \cos u} \,,
\end{equation}
along with $r = a(1-e \cos u)$ and $dt =\sqrt{a^3/Gm} (1-e\cos u) du$.   For the outer binary, we use the fact that 
$dt = \sqrt{P^3/GM} (1+ E \cos F)^{-2} dF$.    Thus the orbit averages may be written
\begin{align}
\left \langle \frac{dY^\alpha}{dt} \right \rangle &= \frac{(1-E^2)^{3/2}}{(2\pi)^2} \int_0^{2\pi} A^\alpha (1-e\cos u) du 
\nonumber \\
& \qquad \times
\int_0^{2\pi} \frac{B^\alpha}{(1+E \cos F)^2} dF \,.
\end{align}

After carrying out the orbital averages, we convert from time $t$ to a dimensionless time scaled by the inner orbital period, namely 
\begin{equation}
\tau \equiv \frac{t}{P_{\rm inner}} = \frac{t}{2\pi} \left( \frac{Gm}{a^3} \right )^{1/2} \,.
\end{equation}
With this scaling, the entire secular dynamics depends on the three dimensionless parameters:
\begin{equation}
\alpha \equiv \frac{m_3}{m} \,, \quad \eta \equiv \frac{m_1m_2}{m^2} \,,  \quad \epsilon \equiv \frac{a}{A}   \,.
\end{equation}
In terms of these parameters, the quantity $\beta = J_b/J_3$ is given by
\begin{equation}
\beta = \eta \frac{(1+\alpha)^{1/2}}{\alpha}  \epsilon^{1/2} \left ( \frac{1-e^2}{1-E^2} \right )^{1/2} \,.
\end{equation}

Through hexadecapole order, we recover the well-known result that $p$, $e$, $P$, and $E$ evolve in such a way that the semimajor axes $a$ and $A$ are constant, in other words
\begin{equation}
\frac{da}{d\tau} = \frac{dA}{d\tau} = 0 \,.
\end{equation}
The secular evolution of the remaining orbit elements is given as follows:

\begin{widetext}
\medskip
\noindent
{\bf Quadrupole order}
\begin{align}
\frac{de}{d\tau} & = \frac{15 \pi}{2} \alpha \epsilon^3  \frac{e(1-e^2)^{1/2}}{(1-E^2)^{3/2}} \sin^2 z \sin \omega \cos \omega \,,
\nonumber \\
\frac{d \iota}{d\tau} &= - \frac{15 \pi}{2} \alpha \epsilon^3  \frac{e^2}{(1-e^2)^{1/2}(1-E^2)^{3/2}} \sin z \cos z  \sin \omega \cos \omega \,,
\nonumber \\
\frac{d\Omega}{d\tau} &= - \frac{3\pi}{2}  \alpha \epsilon^3 \frac{1}{(1-e^2)^{1/2}(1-E^2)^{3/2}} \frac{\sin z \cos z}{\sin \iota} \left (1+4e^2 - 5e^2 \cos^2 \omega \right ) \,,
\nonumber \\
\frac{d\varpi}{d\tau} &= \frac{3\pi}{2}  \alpha \epsilon^3 \frac{(1-e^2)^{1/2}}{(1-E^2)^{3/2}} \left [ 1- \sin^2 z \left ( 4 - 5 \cos^2 \omega \right ) \right ] \,,
\nonumber \\
\frac{dE}{d\tau} & = 0 \,,
\nonumber \\
\frac{d\iota_3}{d\tau} &= - \frac{15 \pi}{2} \eta (1+\alpha)^{1/2}  \epsilon^{7/2}  \frac{e^2}{(1-E^2)^{2}}  \sin z  \sin \omega \cos \omega \,,
\nonumber \\
\frac{d\varpi_3}{d\tau} &= \frac{3\pi}{4}   \eta (1+\alpha)^{1/2}  \epsilon^{7/2} \frac{1}{(1-E^2)^{2}} \left [ 2+3e^2 - 3\sin^2 z \left (1+4e^2 -5e^2 \cos^2 \omega \right ) \right ] \,.
\label{eq2:quadrupole}
\end{align}
At quadrupole order, we recover many of the features of the well-known Kozai-Lidov behavior in hierarchical triple systems, such as the oscillation of $e$ and $\iota$ as the pericenter angle $\omega$ advances.  When $\eta = 0$, the outer orbit is a Keplerian ellipse with constant elements, and the quantity $[a(1-e^2)]^{1/2} \cos z$ is constant under the secular evolution of $e$ and $\iota$; this is proportional to the component of the inner orbit's angular momentum orthogonal to the plane of the third body.   For general $\eta$, there is a fixed point of the orbit elements of the inner orbit ($\dot{e} = \dot{\iota} = \dot{\omega}$), when $\omega = \pi/2$ or $3 \pi/2$, and when $e$ and $z$ satisfy the constraint
\begin{equation}
5\cos^2 z - 3 (1-e^2) + \beta \cos z (1+4e^2) = 0 \,.
\end{equation}
For the outer orbit, the fixed point implies $\dot{\iota}_3 = \dot{E} = 0$, but $\dot{\omega}_3 \ne 0$, in general.

\medskip
\noindent
{\bf Octupole order}
\begin{align}
\frac{de}{d\tau} & = -\frac{15 \pi}{256} \alpha \epsilon^4  \Delta \frac{E(1-e^2)^{1/2}}{(1-E^2)^{5/2}} \biggl(
(4+3e^2) \left [(1+\cos z)(1+10\cos z -15 \cos^2 z) \sin (\omega-\omega_3)
\right .
\nonumber \\
& \qquad
\left .
+(1-\cos z)(1-10\cos z -15 \cos^2 z) \sin (\omega+\omega_3) \right ]
\nonumber \\
& \qquad
-105 e^2 \sin^2 z \left [ (1+\cos z) \sin (3\omega-\omega_3)+(1-\cos z) \sin (3 \omega+\omega_3) \right ] 
 \biggr ) \,,
\nonumber \\
\frac{d \iota}{d\tau} &= - \frac{15 \pi}{256}  \alpha \epsilon^4 \Delta \frac{E e}{(1-e^2)^{1/2}(1-E^2)^{5/2}} \sin z \biggl (
(4+3e^2) \left [(1+10\cos z -15 \cos^2 z) \sin (\omega-\omega_3)
\right .
\nonumber \\
& \qquad
\left .
-(1-10\cos z -15 \cos^2 z) \sin (\omega+\omega_3) \right ]
\nonumber \\
& \qquad
-35 e^2  \left [ (1+\cos z) (1-3\cos z)\sin (3\omega-\omega_3)-(1-\cos z) (1+3\cos z)\sin (3 \omega+\omega_3) \right ]
\biggr ) \,,
\nonumber \\
\frac{d\Omega}{d\tau} &=  \frac{15\pi}{256}  \alpha \epsilon^4 \Delta \frac{Ee}{(1-e^2)^{1/2}(1-E^2)^{5/2}} \frac{\sin z}{\sin \iota} 
\biggl ( 
(4+3e^2) \left [(11-10\cos z -45 \cos^2 z) \cos (\omega-\omega_3)
\right .
\nonumber \\
& \qquad
\left .
-(11+10\cos z -45 \cos^2 z) \cos (\omega+\omega_3) \right ]
\nonumber \\
& \qquad
-35 e^2  \left [ (1+\cos z) (1-3\cos z)\cos (3\omega-\omega_3)-(1-\cos z) (1+3\cos z)\cos (3 \omega+\omega_3) \right ]
\biggr ) \,,
\nonumber \\
\frac{d\varpi}{d\tau} &= -\frac{15\pi}{256}  \alpha \epsilon^4 \Delta \frac{E (1-e^2)^{1/2}}{e (1-E^2)^{5/2}} \biggl (
(4+9e^2) \left [(1+\cos z)(1+10\cos z -15 \cos^2 z) \cos (\omega-\omega_3)
\right .
\nonumber \\
& \qquad
\left .
+(1-\cos z)(1-10\cos z -15 \cos^2 z) \cos (\omega+\omega_3) \right ]
\nonumber \\
& \qquad
-105 e^2 \sin^2 z \left [ (1+\cos z) \cos (3\omega-\omega_3)+(1-\cos z) \cos (3 \omega+\omega_3) \right ] 
\biggr )\,,
\nonumber \\
\frac{dE}{d\tau} & =\frac{15\pi}{256}   \eta (1+\alpha)^{1/2}  \epsilon^{9/2} \Delta \frac{e }{ (1-E^2)^{2}} 
\biggl ( 
(4+3e^2) \left [(1+\cos z)(1+10\cos z -15 \cos^2 z) \sin (\omega-\omega_3)
\right .
\nonumber \\
& \qquad
\left .
-(1-\cos z)(1-10\cos z -15 \cos^2 z) \sin (\omega+\omega_3) \right ]
\nonumber \\
& \qquad
-35 e^2 \sin^2 z \left [ (1+\cos z) \sin (3\omega-\omega_3)-(1-\cos z) \sin (3 \omega+\omega_3) \right ] 
\biggr )
 \,,
\nonumber \\
\frac{d\iota_3}{d\tau} &= \frac{15\pi}{256}  \eta (1+\alpha)^{1/2}  \epsilon^{9/2} \Delta \frac{E e }{(1-E^2)^{3}}   \sin z  
\biggl (
(4+3e^2) \left [(1+10\cos z -15 \cos^2 z) \sin (\omega-\omega_3)
\right .
\nonumber \\
& \qquad
\left .
+ (1-10\cos z -15 \cos^2 z) \sin (\omega+\omega_3) \right ]
\nonumber \\
& \qquad
-35 e^2  \left [ (1+\cos z) (3-\cos z) \sin (3\omega-\omega_3)+(1-\cos z) (3+\cos z)\sin (3 \omega+\omega_3) \right ] 
\biggr ) \,,
\nonumber \\
\frac{d\varpi_3}{d\tau} &=- \frac{15\pi}{256}   \eta (1+\alpha)^{1/2}  \epsilon^{9/2} \Delta \frac{ e (1+4E^2) }{E (1-E^2)^{3}}
\biggl (
(4+3e^2) \left [(1+\cos z) (1+10\cos z -15 \cos^2 z) \cos (\omega-\omega_3)
\right .
\nonumber \\
& \qquad
\left .
+ (1- \cos z) (1-10\cos z -15 \cos^2 z) \cos (\omega+\omega_3) \right ]
\nonumber \\
& \qquad
-35 e^2 \sin^2 z \left [ (1+\cos z) \cos (3\omega-\omega_3)+(1-\cos z) \cos (3 \omega+\omega_3) \right ] 
\biggr ) \,.
\label{eq2:octupole}
\end{align}

It is straightforward to verify that these results are completely equivalent to those of Ford {\em et al.}\ \cite{2000ApJ...535..385F} and NFLRT \cite{2013MNRAS.431.2155N}.  In Appendix \ref{app:dictionary} we provide a dictionary that translates between our osculating orbits language and the Delaunay variables language used in \cite{2000ApJ...535..385F,2013MNRAS.431.2155N}.

\medskip
{\bf Hexadecapole order}
\begin{align}
\frac{de}{d\tau} & = -\frac{315 \pi}{1024} \alpha \epsilon^5 (1-3\eta) \frac{e(1-e^2)^{1/2}}{(1-E^2)^{7/2}}
\nonumber \\
& \quad
\times \biggl( 
(2+3 E^2) \sin^2 z \left [ (4+2e^2) (1-7 \cos^2 z) \sin 2\omega - 21 e^2 \sin^2 z \sin 4\omega \right ]
\nonumber \\
& \qquad
- E^2 \biggl \{ (4+2e^2) \left [(1+\cos z)^2 (1-7\cos z+7\cos^2 z) \sin (2\omega-2\omega_3)
 \right .
 \nonumber \\
& \qquad
\left .
   +(1-\cos z)^2 (1+7\cos z+7\cos^2 z) \sin (2\omega+2\omega_3) \right ]
 \nonumber \\
& \qquad  
   + 21 e^2 \sin^2 z \left [ (1+\cos z)^2  \sin (4\omega-2\omega_3) +(1-\cos z)^2  \sin (4\omega+2\omega_3) \right ] \biggr \}
 \biggr ) \,,
\nonumber \\
\frac{d \iota}{d\tau} &=  \frac{45 \pi}{2048}  \alpha \epsilon^5 (1-3\eta) \frac{\sin z}{(1-e^2)^{1/2}(1-E^2)^{7/2}}  
\nonumber \\
&  \quad
\times \biggl (
14e^2(2+3E^2)\cos z \left [  (4+2e^2) (1-7 \cos^2 z) \sin 2\omega - 21 e^2 \sin^2 z \sin 4\omega \right ]
 \nonumber \\
& \qquad  
+2E^2 (1-7 \cos^2 z)(8+40e^2+15e^4) \sin 2 \omega_3
 \nonumber \\
& \qquad  
+7E^2 e^2 \biggl \{ 4(2+e^2) \left [(1+\cos z) (1-7\cos z+7\cos^2 z) \sin (2\omega-2\omega_3)
 \right .
 \nonumber \\
& \qquad
\left .
   -(1-\cos z) (1+7\cos z+7\cos^2 z) \sin (2\omega+2\omega_3) \right ]
 \nonumber \\
& \qquad
 + 21 e^2 \left [(1-2\cos z) (1+\cos z)^2  \sin (4\omega-2\omega_3) -(1+2\cos z)(1-\cos z)^2  \sin (4\omega+2\omega_3) \right ] \biggr \}
\biggr ) \,,
\nonumber \\
\frac{d\Omega}{d\tau} &=  \frac{45\pi}{2048}  \alpha \epsilon^5 (1-3\eta) \frac{1}{(1-e^2)^{1/2}(1-E^2)^{7/2}} \frac{\sin z}{\sin \iota} 
\nonumber \\
& \quad
\times \biggl ( 
2 (2+3E^2)\cos z \left [  (8+40e^2+15e^4) (3-7 \cos^2 z) 
- 28e^2 (2+e^2) (4-7\cos^2 z) \cos 2\omega +147 e^4 \sin^2 z \cos 4\omega \right ]
 \nonumber \\
& \qquad  
-4 E^2 \cos z (4-7 \cos^2 z)(8+40e^2+15e^4) \cos 2 \omega_3
 \nonumber \\
& \qquad  
+7E^2 e^2 \biggl \{ 2(2+e^2) \left [(1+\cos z) (5+7\cos z-28\cos^2 z) \cos (2\omega-2\omega_3)
 \right .
 \nonumber \\
& \qquad
\left .
   -(1-\cos z) (5-7\cos z-28 \cos^2 z) \cos (2\omega+2\omega_3) \right ]
 \nonumber \\
& \qquad
 - 21 e^2 \left [(1-2\cos z) (1+\cos z)^2  \cos (4\omega-2\omega_3) -(1+2\cos z)(1-\cos z)^2  \cos (4\omega+2\omega_3) \right ] \biggr \}
\biggr ) \,,
\nonumber \\
\frac{d\varpi}{d\tau} &= \frac{45\pi}{1024}  \alpha \epsilon^5 (1-3\eta) \frac{ (1-e^2)^{1/2}}{ (1-E^2)^{7/2}} \nonumber \\
& \quad
\times
\biggl ( 
(2+3E^2) \left [ (4+3e^2)(3-30 \cos^2 z +35 \cos^4 z) - 28 (1+e^2) \sin^2 z (1-7 \cos^2 z) \cos 2 \omega
+147 e^2 \sin^4 z \cos 4 \omega \right ]
\nonumber \\
& \qquad
-10 E^2 (4+3e^2) \sin^2 z (1-7 \cos^2 z) \cos 2 \omega_3
\nonumber \\
& \qquad
+ 7E^2 \biggl \{ 4(1+e^2) \left [ (1+\cos z)^2 (1-7 \cos z +7 \cos^2 z) \cos (2 \omega- 2\omega_3)
 \right .
 \nonumber \\
& \qquad
\left .
+ (1-\cos z)^2 (1+7 \cos z +7 \cos^2 z) \cos(2 \omega + 2\omega_3) \right ]
\nonumber \\
& \qquad
+21 e^2 \sin^2 z \left [ (1+\cos z)^2  \cos(4 \omega- 2\omega_3) +(1-\cos z)^2  \cos(4 \omega+ 2\omega_3) \right ]
\biggr \}
\biggr )\,,
\nonumber \\
\frac{dE}{d\tau} & = -\frac{45\pi}{2048}   \eta (1-3\eta) (1+\alpha)^{1/2}  \epsilon^{11/2}  \frac{E }{ (1-E^2)^{3}} 
\nonumber \\
& \quad
\times
 \biggl ( 
2 (8+40e^2+15e^4) \sin^2 z (1- 7 \cos^2 z)  \sin 2\omega_3
\nonumber \\
& \qquad
+28 e^2 (2+e^2) \left [ (1+\cos z)^2 (1-7 \cos z +7 \cos^2 z) \sin(2 \omega- 2\omega_3)
 \right .
 \nonumber \\
& \qquad
\left .
- (1-\cos z)^2 (1+7 \cos z +7 \cos^2 z) \sin(2 \omega + 2\omega_3) \right ]
\nonumber \\
& \qquad
+147 e^4 \sin^2 z \left [ (1+\cos z)^2  \sin(4 \omega- 2\omega_3) -(1-\cos z)^2  \sin(4 \omega+ 2\omega_3) \right ]
\biggr )
 \,,
\nonumber \\
\frac{d\iota_3}{d\tau} &= \frac{45\pi}{2048}  \eta (1-3\eta) (1+\alpha)^{1/2}  \epsilon^{11/2}  \frac{\sin z }{ (1-E^2)^{4}} 
 \nonumber \\
& \quad
\times \biggl ( 
14e^2(2+3E^2) \left [  (4+2e^2) (1-7 \cos^2 z) \sin 2\omega - 21 e^2 \sin^2 z \sin 4\omega \right ]
 \nonumber \\
& \qquad  
+2E^2 (8+40e^2+15e^4) \cos z (1-7 \cos^2 z) \sin 2 \omega_3
 \nonumber \\
& \qquad  
-7E^2 e^2 \biggl \{ 4(2+e^2) \left [(1+\cos z) (1-7\cos z+7\cos^2 z) \sin (2\omega-2\omega_3)
 \right .
 \nonumber \\
& \qquad
\left .
   + (1-\cos z) (1+7\cos z+7\cos^2 z) \sin (2\omega+2\omega_3) \right ]
 \nonumber \\
& \qquad
 + 21 e^2 \left [(2-\cos z) (1+\cos z)^2  \sin (4\omega-2\omega_3) + (2+ \cos z)(1-\cos z)^2  \sin (4\omega+2\omega_3) \right ] \biggr \}
 \biggr )
 \,,
\nonumber \\
\frac{d\varpi_3}{d\tau} &= \frac{45\pi}{4096}  \eta (1-3\eta) (1+\alpha)^{1/2}  \epsilon^{11/2}  \frac{1 }{ (1-E^2)^{4}} 
\nonumber \\
& \quad
\times \biggl (
(4+3E^2) \left [ (8+40e^2+15e^4)(3-30 \cos^2 z +35 \cos^4 z) 
 \right .
 \nonumber \\
& \qquad
\left .
- 140e^2 (2+e^2) \sin^2 z (1-7 \cos^2 z) \cos 2 \omega
+735 e^4 \sin^4 z \cos 4 \omega \right ]
\nonumber \\
& \qquad
- (2+5 E^2) \biggl \{ 
2(8+40e^2+15e^4)  \sin^2 z (1-7 \cos^2 z) \cos 2 \omega_3
\nonumber \\
& \qquad
- 28 e^2 (2+e^2) \left [ (1+\cos z)^2 (1-7 \cos z +7 \cos^2 z) \cos (2 \omega- 2\omega_3)
 \right .
 \nonumber \\
& \qquad
\left .
+ (1-\cos z)^2 (1+7 \cos z +7 \cos^2 z) \cos(2 \omega + 2\omega_3) \right ]
\nonumber \\
& \qquad
-147  e^4 \sin^2 z \left [ (1+\cos z)^2  \cos(4 \omega- 2\omega_3) +(1-\cos z)^2  \cos(4 \omega+ 2\omega_3) \right ]
\biggr \}
\biggr )
 \,.
 \label{eq2:hexadecapole}
\end{align}
At all three orders, these equations satisfy the three constraints (\ref{eq2:Jtests}) related to the conservation of total angular momentum.

Substituting the definitions (\ref{eq2:keplerorbit1}) and (\ref{eq2:keplerorbit2}) into the expression (\ref{eq2:energy}) for the conserved energy and averaging over time, we obtain the expression
\begin{align}
{\cal E} &= - \frac{Gm^2 \eta}{2a} - \frac{GM^2 \eta_3}{2A} + \frac{GM^2 \eta \, \eta_3 a^2}{8A^3 (1-E^2)^{3/2}} 
\nonumber \\
& \quad
\times \biggl [  1+9 e^2 - 3(1+4e^2) \cos^2 z - 15 e^2 \cos^2 \omega \sin^2 z 
\nonumber \\
& \qquad
+ \frac{15}{64} \frac{a}{A} \frac{eE}{(1-E^2)} \Delta \biggl \{ (4+3e^2) \left [ (1+\cos z)(1+10\cos z -15\cos^2 z) \cos(\omega-\omega_3)
 \right .
 \nonumber \\
& \qquad
\left .
+(1-\cos z)(1-10\cos z -15\cos^2 z) \cos(\omega+\omega_3) \right ]
\nonumber \\
& \qquad
- 35 e^2 \sin^2 z \left [ (1+\cos z) \cos(3 \omega-\omega_3)+(1-\cos z) \cos(3\omega+\omega_3) \right ]
  \biggr \}
\nonumber \\
& \qquad
  -\frac{9}{1024} \left ( \frac{a}{A} \right )^2 \frac{1-3\eta}{(1-E^2)^2} \biggl \{ (2+3E^2) \left [
  (8+40e^2+15e^4) (3-30 \cos^2 z +35 \cos^4 z) 
 \right .
 \nonumber \\
& \qquad
\left .
- 140 e^2 (2+e^2 )  \sin^2 z (1-7 \cos^2 z) \cos 2 \omega +735 e^4 \sin^4 z \cos 4\omega \right ]
 \nonumber \\
& \qquad
- E^2 \biggl (  10(8+40e^2+15e^4) \sin^2 z (1-7 \cos^2 z) \cos 2 \omega_3
 \nonumber \\
& \qquad
-140 e^2 (2+e^2) \left [ (1+\cos z)^2 (1-7 \cos z +7 \cos^2 z) \cos (2 \omega- 2\omega_3)
 \right .
 \nonumber \\
& \qquad
\left .
+ (1-\cos z)^2 (1+7 \cos z +7 \cos^2 z) \cos(2 \omega + 2\omega_3) \right ]
 \nonumber \\
& \qquad
- 735 e^4 \sin^2 z \left [ (1+\cos z)^2  \cos(4 \omega- 2\omega_3) +(1-\cos z)^2  \cos(4 \omega+ 2\omega_3) \right ]
\biggr )
\biggr \}
   \biggr ]\,.
   \label{eq2:energyaverage}
\end{align}
\end{widetext}
The quadrupole and octupole contributions agree with the corresponding contributions to the ``energy function'' $F$, displayed in Eqs.\ (8) - (11) of \cite{2011ApJ...742...94L}; in that calculation, $\eta$ was chosen to vanish, the constant pericenter of the outer orbit was chosen to lie along the $X$-axis, and thus $\omega_3  = \pi  -\Omega$.

\section{Case studies of the effects of hexadecapole contributions}
\label{sec:numerical}

We now turn to the numerical analysis of the secular evolution of the orbital elements for cases of astrophysical interest.   The two semimajor axes $a$ and $A$ are constants of the motion.  The precession of the nodal angle $d\Omega/d\tau$ is of no internal dynamical interest; it represents an irrelevant rotation of the entire system about the conserved total angular momentum vector.   None of the evolution equations depends on $\Omega$.  The equation for
$d\Omega/d\tau$ is useful only for constructing the equations of evolution for $\omega$ and $\omega_3$ using the relations $\dot{\omega} = \dot{\varpi} - \dot{\Omega} \cos \iota$ and 
$\dot{\omega}_3 = \dot{\varpi}_3 - \dot{\Omega} \cos \iota_3$.  The individual inclinations $\iota$ and $\iota_3$ can be directly linked  to the relative inclination angle $z$ via Eq.\ (\ref{eq2:inclinations}), and only $z$ appears in the equations.   Thus the dynamical system reduces to five evolution equations for the five variables $e$, $E$, $z$, $\omega$ and $\omega_3$, depending only on the three dimensionless parameters $\alpha = m_3/m$, $\eta = m_1 m_2/m^2$ and $\epsilon = a/A$.   The only place where the actual mass or distance scale enters is in the conversion from the dimensionless time $\tau$ to real time $t$ via the scaling $t = P_{\rm inner} \tau = 2 \pi \tau (a^3/Gm)^{1/2}$.

The foregoing remarks apply only in Newtonian gravity.   In the real world, general relativity should be included, and indeed it is well known that the simplest quadrupole-order Kozai-Lidov oscillations can be strongly suppressed if the rate of relativistic advance of the pericenter of the inner binary is large enough \cite{1997Natur.386..254H}.   Including the leading contribution of general relativity forces us to introduce an additional dimensionless parameter $\delta$ to the problem, given by
\begin{equation}
\delta \equiv \frac{Gm}{c^2 a} = 9.8736 \times 10^{-9} \left (\frac{m}{M_\odot} \right ) \left (\frac{{\rm a.u.}}{a} \right )\,,
\end{equation}
where $c$ is the speed of light.   The dominant effect is to add to the pericenter advances of the two orbits the terms
\begin{align}
\frac{d\varpi}{d\tau} &= 6\pi \frac{\delta}{1-e^2} \,,
\nonumber\\
\frac{d\varpi_3}{d\tau} &= 6\pi \frac{\delta(1+\alpha) \epsilon}{1-E^2} \,.
\end{align}  
Additional relativistic effects, such as those studied in \cite{2013ApJ...773..187N}, will be the subject of future work.

With three fundamental parameters (four if we include general relativity) and 5 dynamical variables, a complete exploration of the full parameter space is beyond the scope of this paper.  Instead we will analyze the effects of the hexadecapole contribution on a selection of case studies that have appeared in the literature.   Most of these have been presented by Naoz and collaborators \cite{2011Natur.473..187N,2013MNRAS.431.2155N}, who first pointed out examples where orbital flips and excursions to very large eccentricities induced by octupole-order terms were astrophysically interesting.   We will find that, in almost all cases, the hexadecapole and GR contributions make only small quantitative differences, but do not impact the orbital-flip or large-eccentricity phenomena.   

Table \ref{table:params} lists the specific parameters and initial conditions for the cases studied in this section.

\subsection{Hot Jupiters}

\begin{figure}[t]
\begin{center}

\includegraphics[width=3.4in]{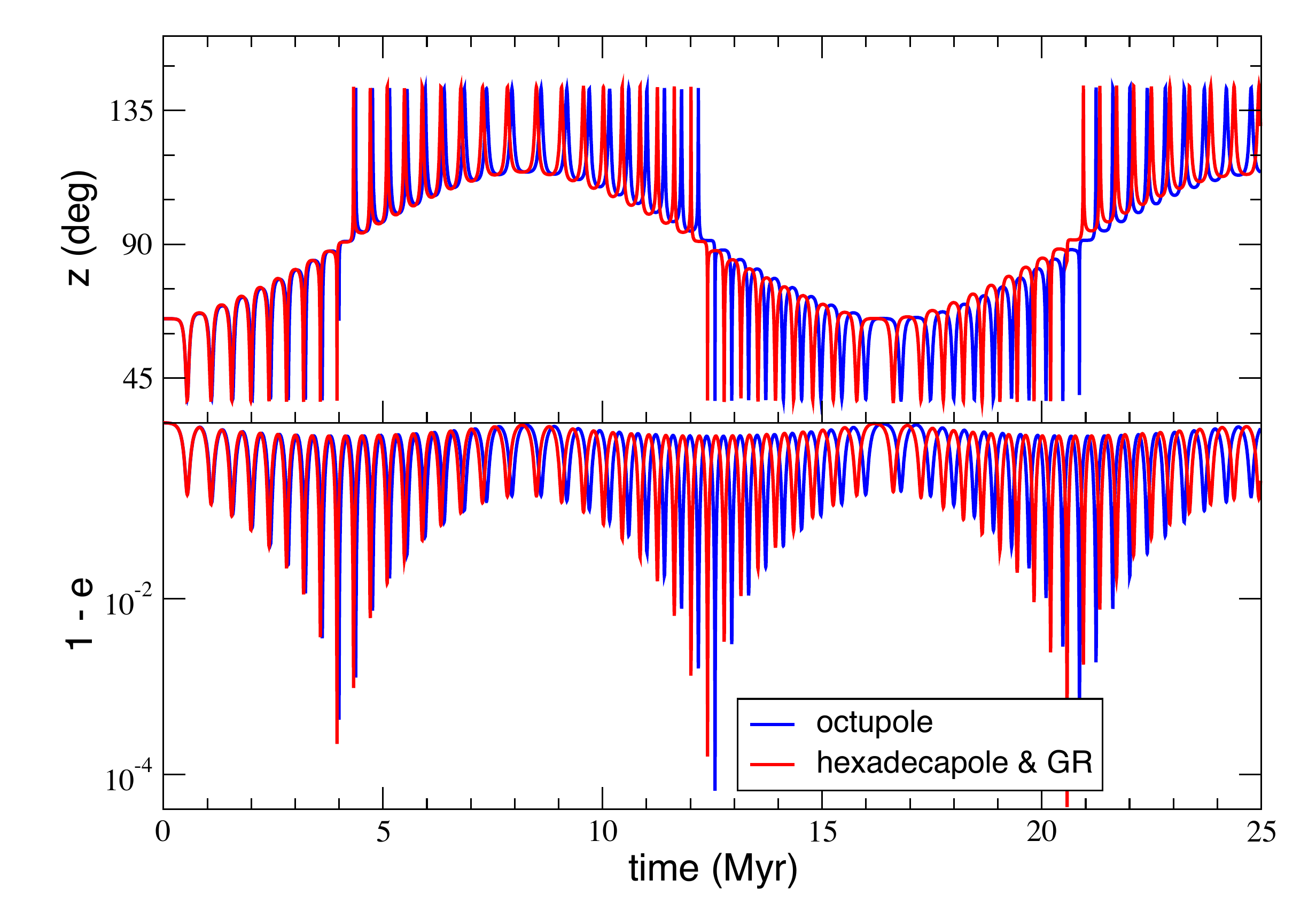}

\caption{Orbital flips and eccentricity excursions in a Jupiter-Sun system perturbed by a distant brown dwarf.   Blue: octupole order, without GR (equivalent to Fig.\ 3  of \cite{2013MNRAS.431.2155N}).  Red: hexadecapole order with GR. Parameters and initial orbit elements are listed in Table \ref{table:params}.  (Color figures in online version.)
\label{fig:NaozMNRASFig3_zcombined} }
\end{center}
\end{figure}

In their seminal discussion of the possibility of hot Jupiters in retrograde orbits, NFLRT considered an inner binary of a Jupiter orbiting a solar-mass star with $a=6$  a.u., perturbed by a brown-dwarf star with a mass of $40 M_J$ and $A = 100$ a.u.    In this case, with $M_\odot = 1047 M_J$, the parameters (including the GR parameter)  take the values
\begin{align}
\alpha &= 0.0382 \, \quad \epsilon = 0.06 \,, \quad  \eta = 9.53 \times 10^{-4}  \,, 
\nonumber \\
\quad \Delta &= 0.998 \,, \quad \delta = 1.63 \times 10^{-9} \,.
\end{align}
The initial conditions chosen by NFLRT were
\begin{equation}
e =  0.001 \,, \, E = 0.6 \,, \, z = 65^{\rm o} \,, \, \omega = 45^{\rm o} \,, \, \omega_3 = 0^{\rm o} \,.
\end{equation}
We evolve the secular planetary equations for $1.7 \times 10^6$ orbits of the inner binary (corresponding to about $2.5 \times 10^7$ years) for four cases, octupole order, with and without GR precessions and  hexadecapole order, with and without GR precessions.   The four cases yield very similar results and so we show only two of the cases.  Figure \ref{fig:NaozMNRASFig3_zcombined} 
shows the inclination angle $z$  and $\log(1-e)$ against time.   Plotted in blue is the Newtonian evolution at octupole order, matching very well the results of  \cite{2011Natur.473..187N,2013MNRAS.431.2155N}.  Initially the system undergoes Kozai-Lidov type oscillations in $z$ but with the maximum value of $z$ rising steadily; when $z$ reaches $90^{\rm o}$, the orbit becomes retrograde and the oscillations ``flip''.  Later the orbit flips back to prograde, and so on.     Plotted in red is the full evolution including hexadecapole terms and the GR pericenter precessions.   The pattern of flips and the excursions to large eccentricity are essentially the same as in the octupole case; only the timescale has been shortened slightly, in agreement with the $N$-body integrations carried out by Naoz {\em et al.}\ and shown in their Fig.\ 3 \cite{2013MNRAS.431.2155N}.
In this case, the hexadecapole and relativistic terms do not change the behavior to any significant degree.  

We remark that Carvalho {\em et al.}\ \cite{2016CeMDA.124...73C} found that hexadecapole contributions, derived assuming a fixed third-body orbit, appeared to produce somewhat anomalous flip behavior (their Fig.\ 8), only to be restored to behavior consistent with direct numerical integrations by the dotriocontupole terms, derived under the same assumption (their Fig.\ 9).  In our approach, both orbits are perturbed consistently, and the hexadecapole order results are fully compatible with the $N$-body integrations of \cite{2013MNRAS.431.2155N}.

  \begin{widetext}

\begin{table}[t]
\begin{center}
\caption{Physical parameters and initial conditions for selected case studies} 
\begin{tabular}{l@{\hskip 0.5cm}c@{\hskip 0.5cm}c@{\hskip 0.5cm}c@{\hskip 0.5cm}c@{\hskip 0.5cm}c@{\hskip 0.5cm}c@{\hskip 0.5cm}c@{\hskip 0.5cm}c@{\hskip 0.5cm}c@{\hskip 0.5cm}c}
\hline
System&$m_1$&$m_2$&$m_3$&$a$ (a.u.)&$A (a.u.)$&$e$&$E$&$z$&$\omega$&$\omega_3$ \\
\hline 
Hot Jupiters&$M_J$&$M_\odot$&$40\, M_J$&6&100&0.001&0.6&65&45&0\\
Coplanar Flips&$M_J$&$M_\odot$&$0.03\, M_\odot$&4&50&0.8&0.6&5&0&0\\
%
%
Asteroid-Jupiter&$0$&$M_\odot$&$M_J$&2&5&0.2&0.05&65&0&0\\
Triple star&$0.25 \,M_\odot$&$M_\odot$&$0.6 \,M_\odot$&60&800&0.01&0.6&98&0&0\\
CH Cygni&$0.5 \,M_\odot$&$3.51 \,M_\odot$&$0.909 \,M_\odot$&0.05&0.21&0.32&0.6&72&145&0\\
\hline
\end{tabular}
\label{table:params}
\end{center}
\end{table}

\end{widetext}

\subsection{Orbital flips from nearly coplanar orbits}
\label{sec:coplanar}

\begin{figure}
\begin{center}

\includegraphics[width=3.4in]{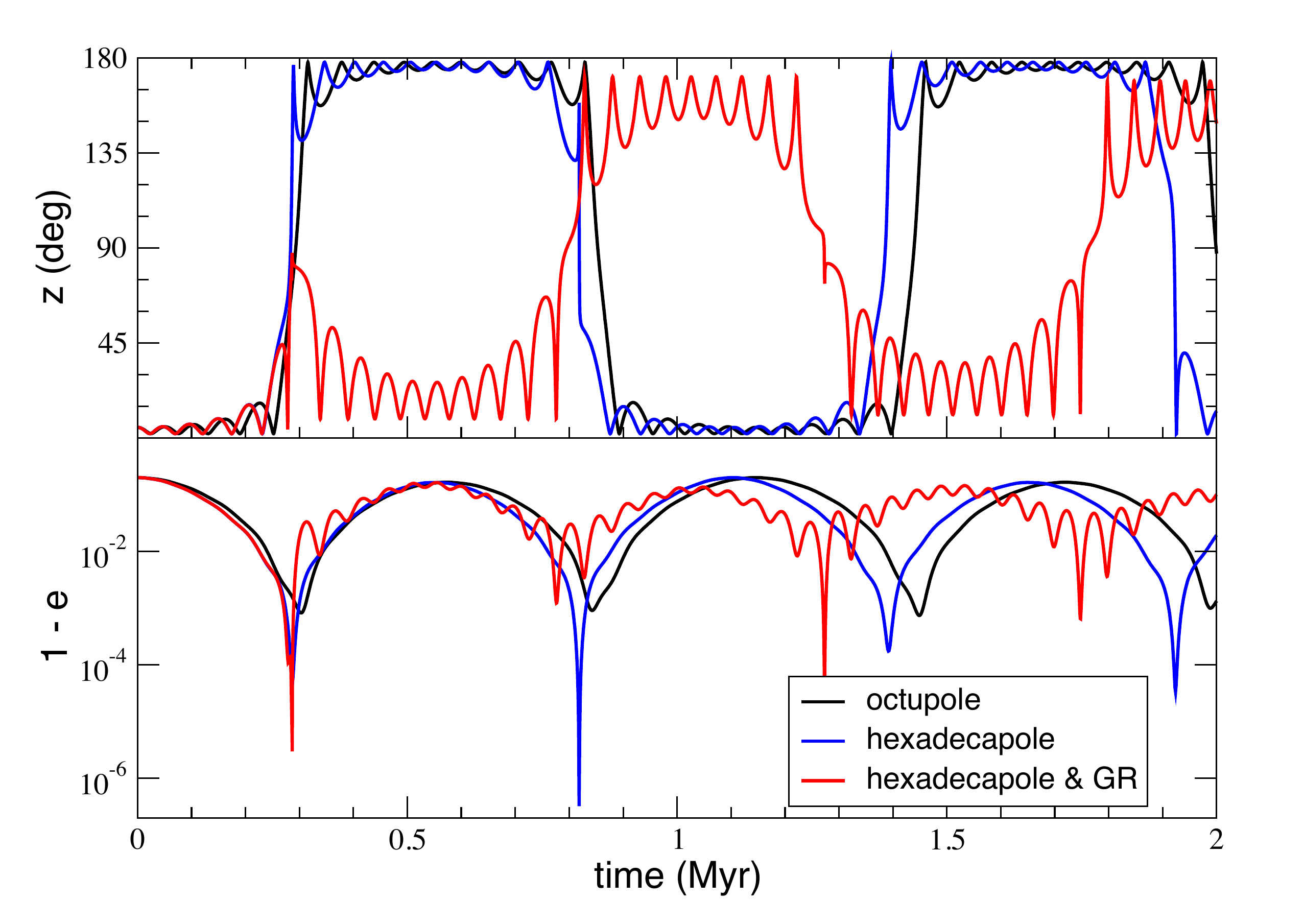}

\caption{Orbital flips and eccentricity excursions in a nearly co-planar Jupiter system.     Black: octupole order. Blue: hexadecapole order. Red: hexadecapole order with GR.  Parameters and initial orbit elements are listed in Table \ref{table:params}. (Color figures in online version.)
\label{fig:LiNaoz} }
\end{center}
\end{figure}

Li {\em et al.}~\cite{2014ApJ...785..116L} discovered the possibility of generating orbital flips and large eccentricities from initially nearly coplanar orbits using the octupole-order equations.  
The inner system was again a Jupiter-Sun binary with $a=4$ a.u., perturbed by a brown dwarf, with $m_3 = 0.03 \, M_\odot$ and $A = 50$ a.u.   The parameters then have the values
 \begin{align}
\alpha &= 0.030 \, \quad \epsilon = 0.08 \,, \quad  \eta = 9.53 \times 10^{-4}  \,, 
\nonumber \\
\quad \Delta &= 0.998 \,, \quad \delta = 1.63 \times 10^{-9} \,,
\end{align}
and the initial conditions are
\begin{equation}
e =  0.8 \,, \, E = 0.6 \,, \, z = 5^{\rm o} \,, \, \omega = 0^{\rm o} \,, \, \omega_3 = 0^{\rm o} \,.
\end{equation}
We evolve the equations for $2.5 \times 10^5$ inner orbits ($2 \times 10^6$ years), for three cases: octupole and hexadecapole orders without GR precessions, and hexadecapole order with  GR precessions.     The results are shown in Fig.\ \ref{fig:LiNaoz}.
At octupole order without GR (upper panel, plotted in black), the system oscillates about small values of $z$ for a while, then migrates quickly to a retrograde orbit, oscillates about the new values for a while, then migrates back.  During the transition the eccentricity reaches extreme values close to unity (lower panel).  Including the hexadecapole terms shortens the timescale slightly (plotted in blue), but otherwise preserves the basic behavior.   These curves are in excellent agreement with results obtained by Li \cite{Liprivate} and by Hamers \cite{Hamersprivate} using both $N$-body codes and orbit element codes to the same multipolar order.   However, including the GR precessions with the hexadecapole terms causes the first flip to abort (plotted in red); subsequent flips are then out of phase with those where GR is not included.   It is evident that the transition from prograde to retrograde orbits is very sensitive to the phases of the two pericenter angles, $\omega$ and $\omega_3$ as the inclination angle $z$ approaches $90^{\rm o}$.  The cumulative precessions in these angles induced by general relativity can turn a transition to retrograde into a bounce back to prograde, and vice versa.

\begin{figure}
\begin{center}

\includegraphics[width=3.4in]{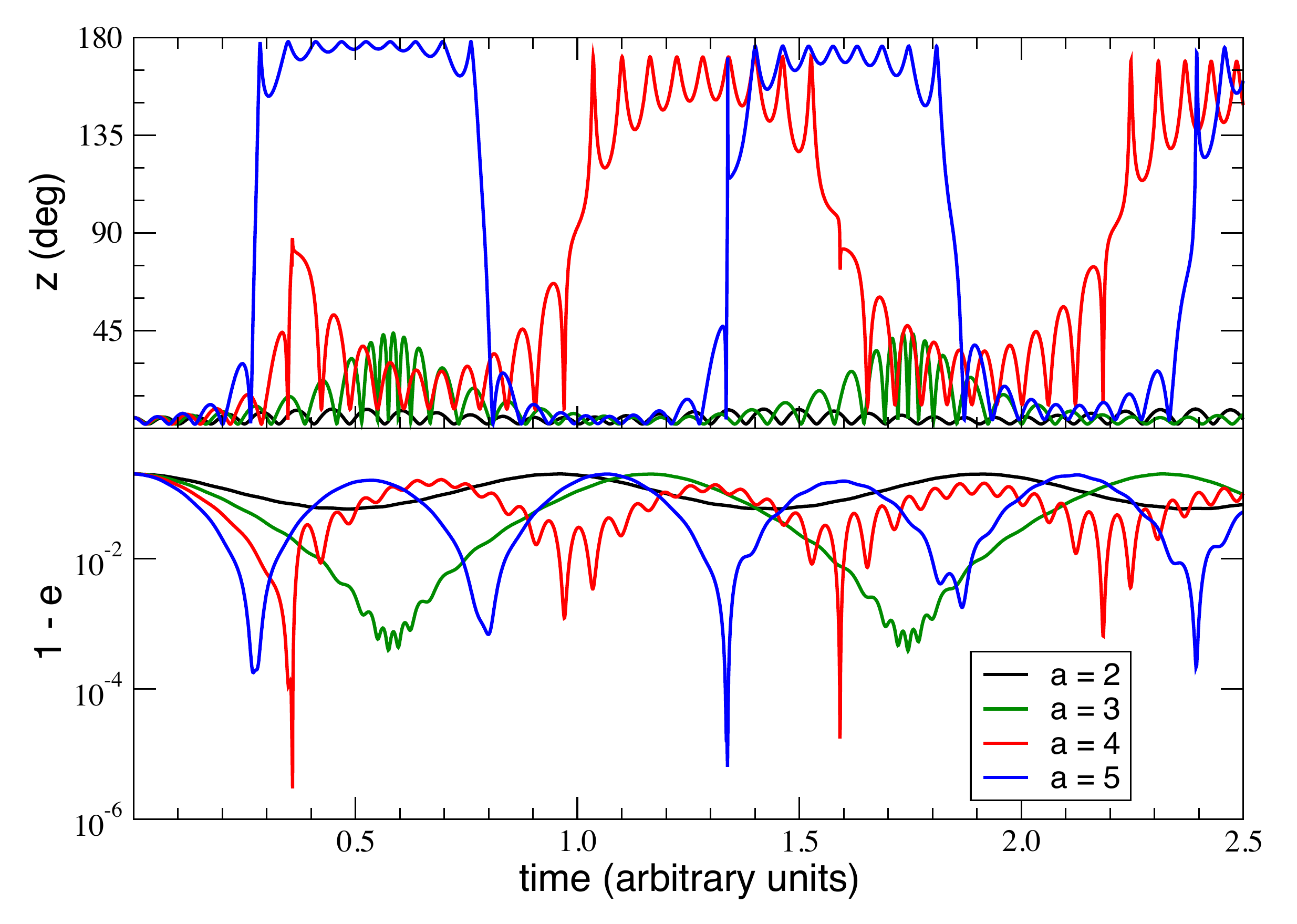}

\caption{Effects of GR on orbital flips and eccentricity excursions in a nearly co-planar Jupiter system.      Blue: $a=5$ a.u.  Red: $a = 4$ a.u. (same as in Fig.\ \ref{fig:LiNaoz}). Green: $a = 3$ a.u. Black: $a = 2$ a.u.
Other parameters and initial orbit elements are listed in Table \ref{table:params}. (Color figures in online version.)
\label{fig:LiNaozGR} }
\end{center}
\end{figure}

We now vary the semimajor axis $a$ of the inner orbit in order to assess the effects of GR.  Holding the other parameters and initial conditions fixed, we obtain the curves shown in Fig.\ \ref{fig:LiNaozGR}.  Here the time scales as $\tau (a/4)^3$, where $a$ is in astronomical units; this timescale is chosen so that similar numbers of Kozai cycles can appear on one plot.   For $a=5$ (blue), the pattern of flips is very similar to that obtained without GR, shown in blue in Fig.\ \ref{fig:LiNaoz}.   For $a=4$ (red), the curve is the same as that shown in red in Fig.\ \ref{fig:LiNaoz}.   For $a = 3$ (green) the migration to large inclinations is suppressed by the more rapid GR precessions, although migrations to large eccentricities still occur.  Finally, for $a=2$, (black) the GR precessions permit only small Kozai-like oscillations about the initial values of $z$ and $e$.  

\begin{figure}
\begin{center}

\includegraphics[width=3.4in]{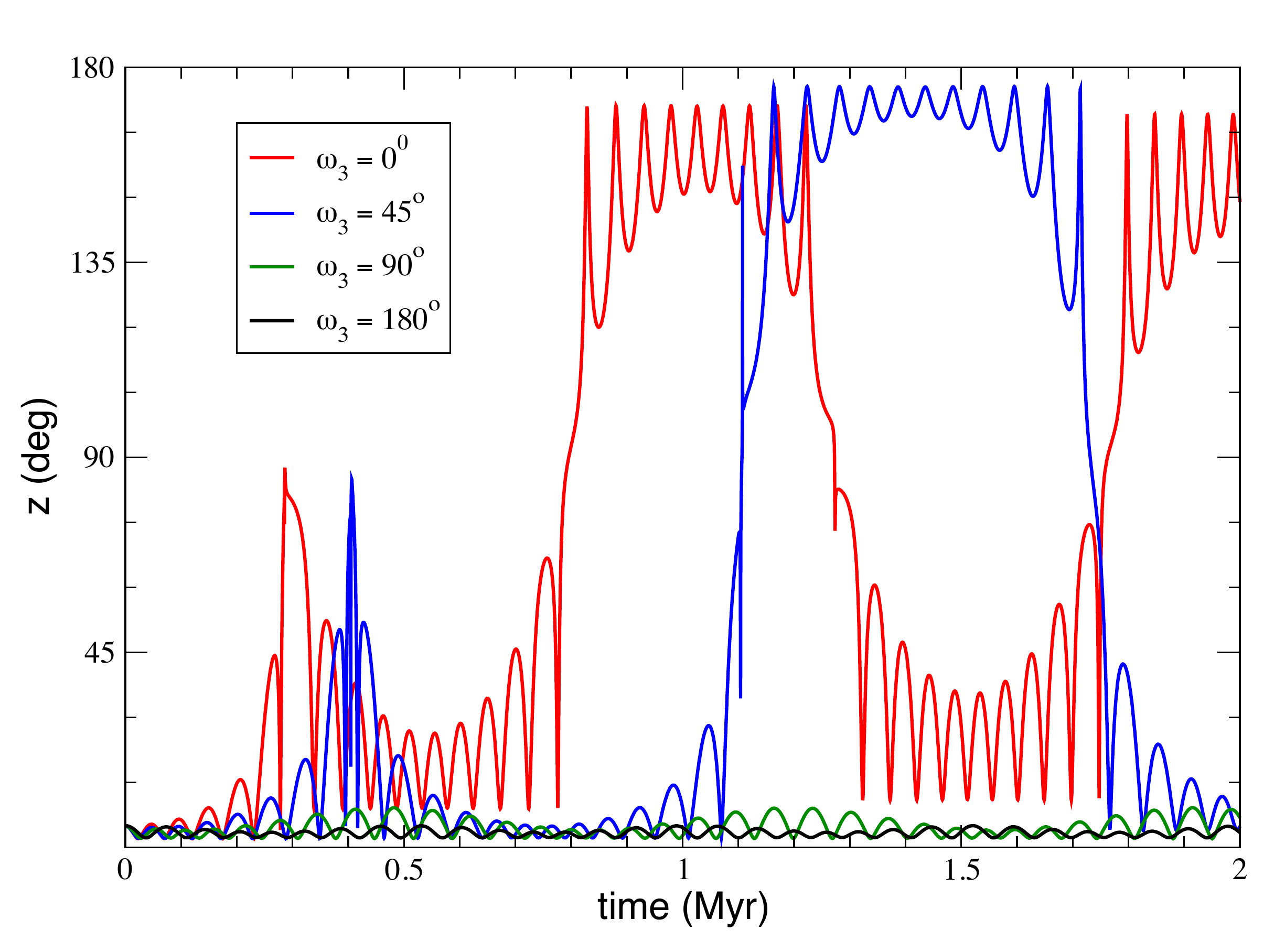}

\caption{Effects of initial pericenter angle on orbital flips in a nearly co-planar Jupiter system for $a =4$ a.u.      Red: $\omega_3 = 0^{\rm o}$ (same as in Fig.\ \ref{fig:LiNaoz}). Blue: $\omega_3 = 45^{\rm o}$.  Green: $\omega_3 = 90^{\rm o}$.   Black: $\omega_3 = 180^{\rm o}$.
Other parameters and initial orbit elements are listed in Table \ref{table:params}. (Color figures in online version.)
\label{fig:LiNaozomega3} }
\end{center}
\end{figure}

For the nominal value $a = 4$ a.u., we also show the sensitivity of orbital flips to the pericenter angles.  Figure \ref{fig:LiNaozomega3} shows evolutions for four initial pericenter angles of body 3: $0^{\rm o}$ (red, same as in Fig.\ \ref{fig:LiNaoz}), $45^{\rm o}$ (blue),  $90^{\rm o}$ (green) and $180^{\rm o}$ (black).   Notice that the initial values $\omega =0^{\rm o}$, $\omega_3 = 0^{\rm o}$ correspond to orbits with the initial pericenters pointing in opposite directions (Fig.\ \ref{fig:orbits}), while  the values $\omega =0^{\rm o}$, $\omega_3 = 180^{\rm o}$ correspond to initial orbits with the pericenters pointing in the same direction.  This dependence is in agreement with results from $N$-body integrations by Li \cite{Liprivate}.  

\subsection{An asteroid Jupiter system}

\begin{figure}[t]
\begin{center}

\includegraphics[width=3.4in]{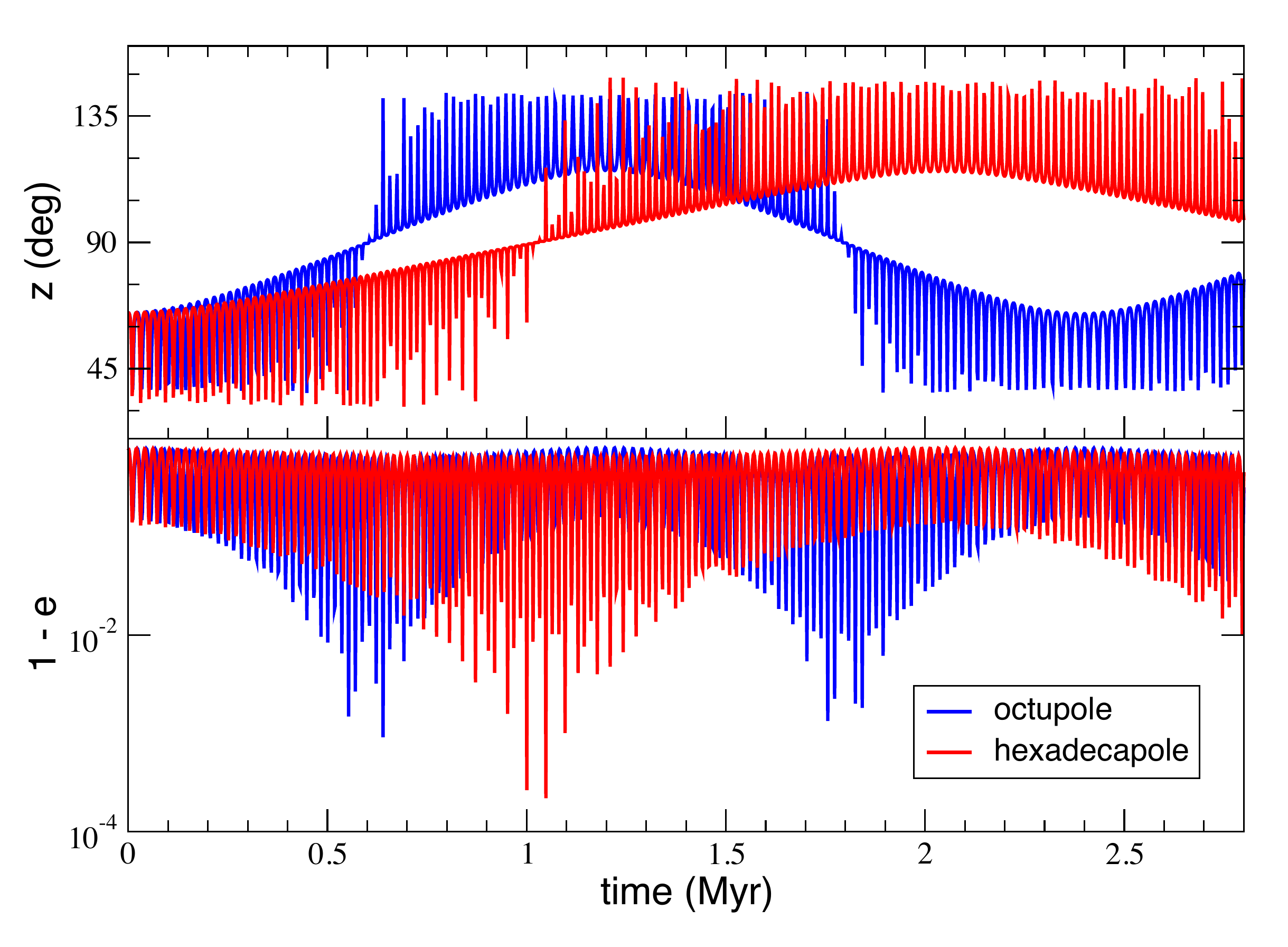}

\caption{Evolution of inclination and eccentricity in an asteroid-Jupiter system.  Blue: octupole order. Red: hexadecapole order.  GR precessions are included. (Color figures in online version.)
 \label{fig:NaozMNRASFig8_combined} }
\end{center}
\end{figure}

NFLRT showed that octupole perturbations could induce orbital flips in a Sun-asteroid-Jupiter system.   In this case, $a = 2$ a.u.\ and $A= 5$ a.u., and the parameters are
 \begin{align}
\alpha &= 9.55 \times 10^{-4} \, \quad \epsilon = 0.4 \,, \quad  \eta = 0  \,, 
\nonumber \\
\quad \Delta &= 1 \,, \quad \delta = 4.92 \times 10^{-9} \,. 
\end{align}
The initial conditions are
\begin{equation}
e =  0.2 \,, \, E = 0.05 \,, \, z = 65^{\rm o} \,, \, \omega = 0^{\rm o} \,, \, \omega_3 = 0^{\rm o} \,.
\end{equation}
We evolve the planetary equations for $10^6$ orbital periods, corresponding to about 2.8 million years, with and without hexadecapole terms.  We include the GR precessions, but they turn out to have negligible effect in this example.   Figure \ref{fig:NaozMNRASFig8_combined} shows the resulting evolutions of $z$ and $e$.   Including the hexadecapole terms stretches the timescale somewhat, in agreement with the full $N$-body numerical evolutions carried out by NFLRT (see Fig.\ 8 of NFLRT).   As in the previous example, the initial choice $\omega_3 = 180^{\rm o}$ leads to no orbital flips. 

\subsection{A triple-star hierarchical system}

\begin{figure}[t]
\begin{center}

\includegraphics[width=3.4in]{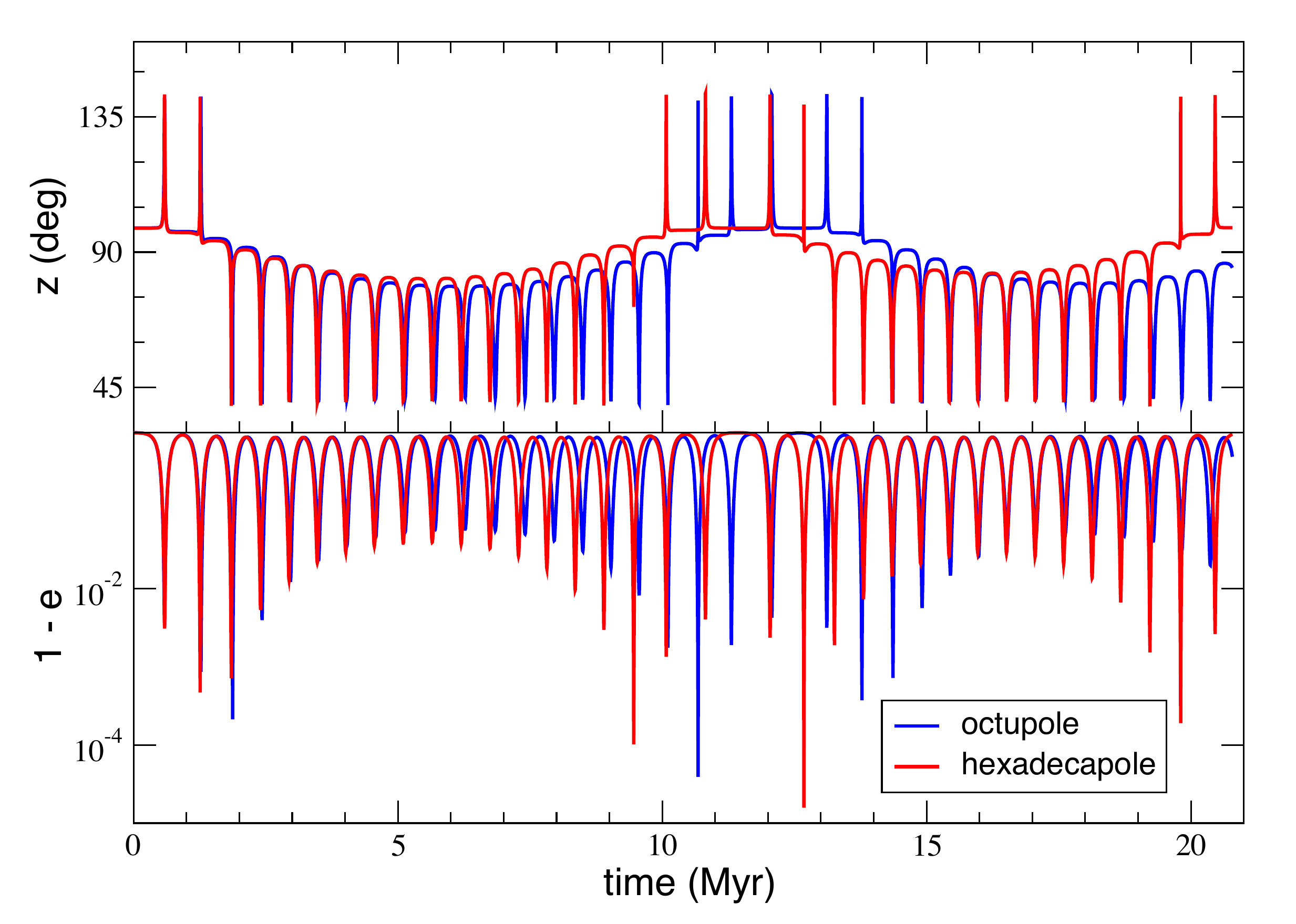}

\caption{Evolution of inclination and eccentricity in a triple-star system.  Blue: octupole order.  Red: hexcadapole order.  (Color figures in online version.)
\label{fig:NaozMNRASFig9} }
\end{center}
\end{figure}

Analyzing a set of hierarchical triple-star systems studied by Fabrycky and Tremaine \cite{2007ApJ...669.1298F}, NFLRT again found orbital flip behavior (Fig.\ 9 of \cite{2013MNRAS.431.2155N}).   The system studied consists of an inner binary with $m_1  = 0.25 M_\odot$, $m_2 = M_\odot$, $a= 60$ a.u., and an outer star, with $m_3 = 0.6 M_\odot$, $a= 800$ a.u.  In this example, the parameters are
 \begin{align}
\alpha &= 0.48 \, \quad \epsilon = 0.075 \,, \quad  \eta = 0.16  \,, 
\nonumber \\
\quad \Delta &= 0.6 \,, \quad \delta = 2.05 \times 10^{-10} \,. 
\end{align}
The initial conditions are
\begin{equation}
e =  0.01 \,, \, E = 0.6 \,, \, z = 98^{\rm o} \,, \, \omega = 0^{\rm o} \,, \, \omega_3 = 0^{\rm o} \,.
\end{equation}
In this example, the initial inner orbit is already retrograde.
We evolve for $5 \times 10^4$ orbits, corresponding to $2 \times 10^7$ years.   The results are shown in Fig.\ \ref{fig:NaozMNRASFig9}.   The octupole-order curves (blue) agree well with the curves displayed in Fig.\ 9 of \cite{2013MNRAS.431.2155N}, while the hexadecapole contributions (red) preserve the flips with minor changes.   However, if we include the hexadecapole orders and decrease $a$, making the inner binary more relativistic, while holding the other parameters and initial conditions fixed, then the flips to prograde become progressively more sporadic, finally disappearing completely when $a = 18$ a.u.

\subsection{The CH Cygni system}

\begin{figure}[t]
\begin{center}

\includegraphics[width=3.4in]{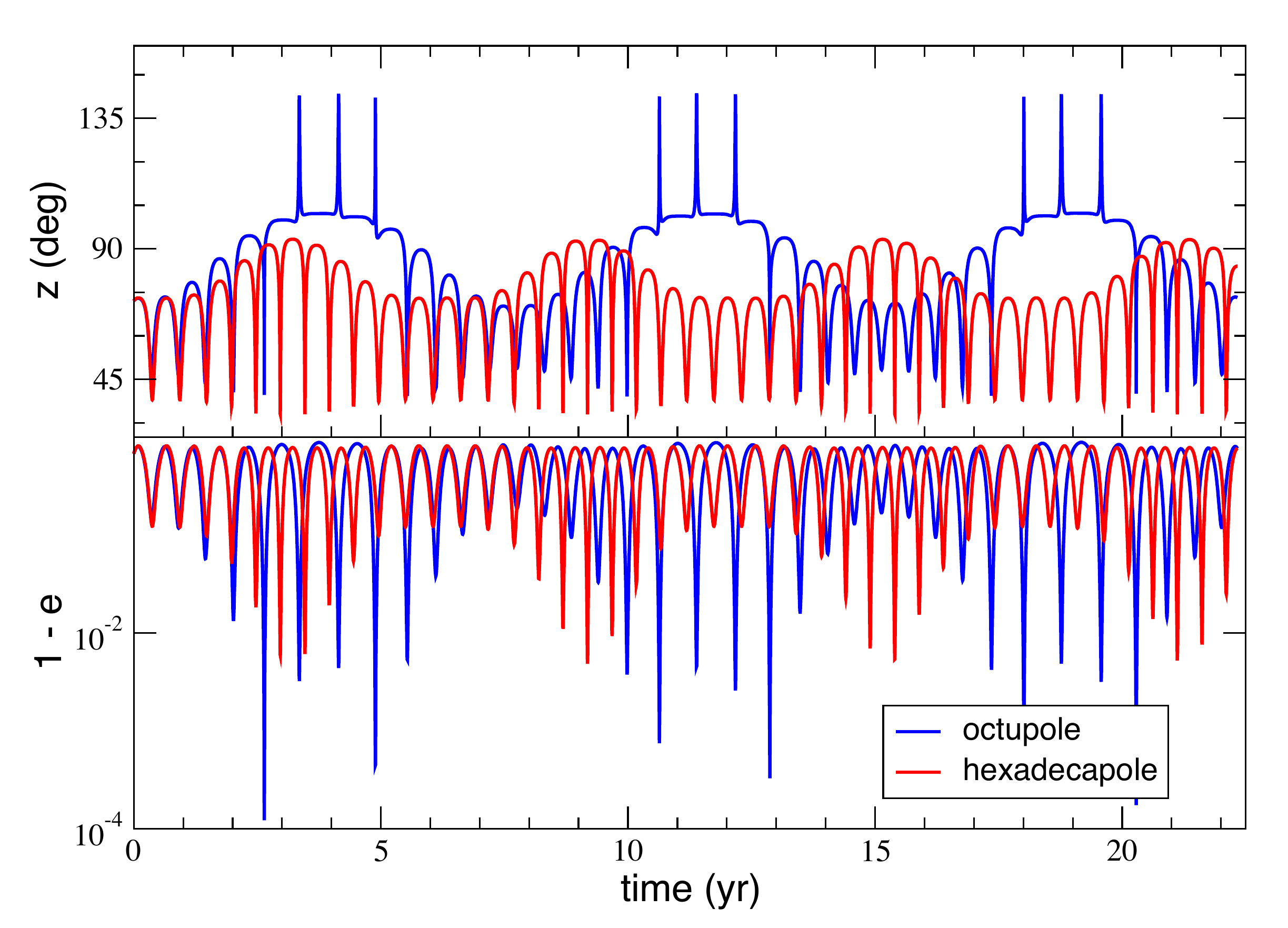}        

\caption{Evolution of $z$ and $e$ in the CH Cygni system. Blue: octupole order.  Red: hexcadapole order.  GR precessions are included.  (Color figures in online version.) \label{fig:NaozMNRASFig11} }
\end{center}
\end{figure}

Using the best fit parameters for the triple system CH Cygni from Mikkola and Tanikawa \cite{1998AJ....116..444M}, NFLRT showed that including the octupole-order contributions changed the evolution from conventional Kozai oscillations to orbital flips and excursions to large eccentricity.  
The parameters are
 \begin{align}
\alpha &= 0.227 \,, \quad \epsilon = 0.238 \,, \quad  \eta = 0.109  \,, 
\nonumber \\
\quad \Delta &= 0.751 \,, \quad \delta = 7.89 \times 10^{-7} \,, 
\end{align}
and the initial conditions are
\begin{equation}
e =  0.32 \,, \, E = 0.6 \,, \, z = 72^{\rm o} \,, \, \omega = 145^{\rm o} \,, \, \omega_3 = 0^{\rm o} \,.
\end{equation}
We evolve for 4000 orbits, corresponding to about 22 years, with results shown in  Fig.\ \ref{fig:NaozMNRASFig11}.  The octupole-order results (blue) closely match those of \cite{2013MNRAS.431.2155N}, Fig.\ 11, showing both orbital flips and large eccentricity excursions.  But in this case, with hexadecapole contributions (red), the flips are suppressed and the eccentricity excursions are reduced.   GR precessions were included, but make no discernable difference in this example.   

\section{Equal-mass inner binaries}
\label{sec:equalmass}

When the bodies making up the inner binary have equal masses, the octupole terms vanish, leaving only the quadrupole and hexadecapole contributions.   It is therefore interesting to explore whether the hexadecapole terms alone can generate orbital flips and large eccentricities.   Since precise equality of masses is rare, this special case might not be of generic astrophysical interest, although it might be relevant for inner binaries consisting of neutron stars, whose masses tend to cluster around $1.4 M_\odot$. 

In the equal-mass case, $\eta = 1/4$ and $\Delta = 0$, and thus the  free parameters reduce to three: $\alpha$, $\epsilon$ and the GR parameter $\delta$.   Because the hexadecapole terms are smaller than the quadrupole terms by a factor $\epsilon^2$, then if $\epsilon$ is too small, hexadecapole effects are too small to be of any consequence.      One ``sweet spot'', where orbital flips can be induced by hexadecapole terms alone occurs around the values $\epsilon \sim 0.1$ and $\alpha \sim 10$.   Note that the combination $\alpha \epsilon^3$, which controls the leading quadrupole effects, is still small; this constraint must hold so that the problem remains within the perturbative regime.   

\begin{figure}[t]
\begin{center}

\includegraphics[width=3.4in]{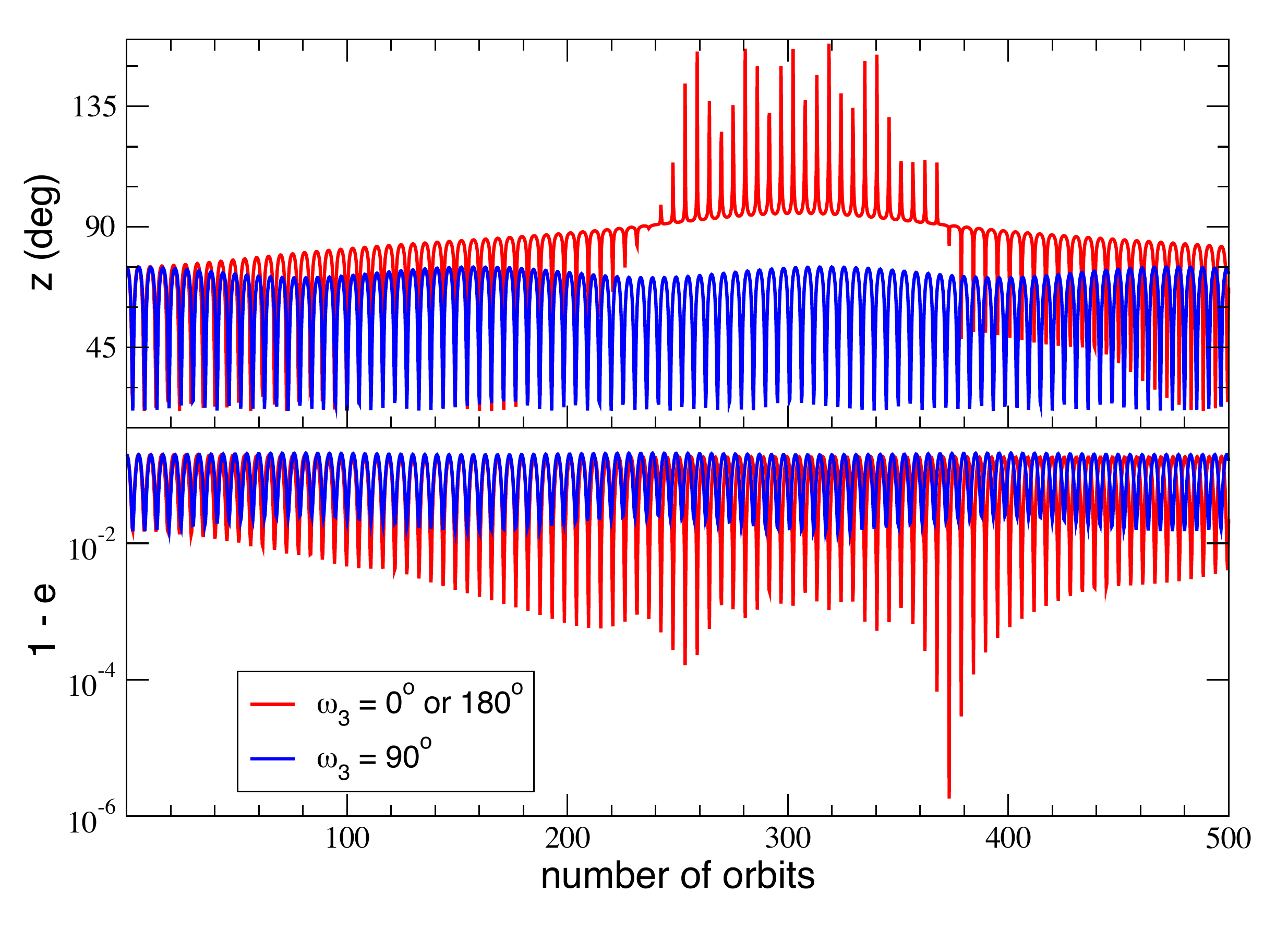}       

\caption{Evolution of $z$ and $e$ for an equal-mass inner binary. Red: $\omega_3 = 0^{\rm o}$ or $180^{\rm o}$.  Blue: $\omega_3 = 90^{\rm o}$.  GR precessions are included.  (Color figures in online version.) \label{fig:EqualMass} }
\end{center}
\end{figure}

The first example is displayed in Fig.\ \ref{fig:EqualMass}.   The chosen parameters are:
\begin{align}
\alpha &= 10.714 \,, \quad \epsilon = 0.127 \,, \quad   \delta = 3.95 \times 10^{-9} \,. 
\label{eq3:params1}
\end{align}
The initial conditions are
\begin{align}
e &=  0.8 \,, \, E = 0.6 \,, \, z = 75^{\rm o} \,, \, \omega = 0^{\rm o} \,, 
\nonumber \\
\omega_3 &= 0^{\rm o} \, {\rm or} \, 180^{\rm o} \, {\rm (red)} \,, \, 90^{\rm o} \, {\rm (blue)}\,.
\end{align}
A specific system with these parameters consists of two $1.4 \, M_\odot$ neutron stars orbiting a $30 \, M_\odot$ star or black hole, with $a = 7$ a.u. and $A = 55$ a.u.    Scaling all masses and semimajor axes by a common factor $\zeta$ yields identical evolutions, since the three parameters of Eq.\ (\ref{eq3:params1}) are unchanged. Only the timescale set by the inner orbital period changes, scaling by $\zeta$.   Evolving the system for 500 orbits of the inner binary, we find that the evolution for $\omega_3 = 0^{\rm o}$ (initial pericenters in opposite directions along the line of nodes) is identical to that for $\omega_3 = 180^{\rm o}$ (initial pericenters in the same direction), resulting in an orbital flip (red curves in Fig.\ \ref{fig:EqualMass}), while the evolution for $\omega_3 = 90^{\rm o}$ does not show flips.   This is in contrast to the cases where octupole terms dominate, where $\omega_3 = 0^{\rm o}$ leads to flips while $\omega_3 = 180^{\rm o}$ does not.   This makes sense because, as can be seen from Eqs.\ (\ref{eq2:octupole}), the octupole terms change sign under the transformation $\omega_3 \to \omega_3 + \pi$, whereas the hexadecapole terms  in Eqs.\ (\ref{eq2:hexadecapole}) are invariant under that transformation.   On the other hand, many pieces of the hexadecapole terms change sign under the transformation $\omega_3 \to \omega_3 + \pi/2$, and as a consequence, the initial angle $\omega_3 = 90^{\rm o}$ yields no flips.  

In the foregoing example, the evolutions are the same whether the GR precessions are included or not.   We can investigate when GR effects become important by ``dialing up'' the GR parameter $\delta$ while holding $\alpha$ and $\epsilon$ fixed.  This is equivalent either to reducing $a$ and $A$ by the same factor, holding the masses fixed, or to increasing all the masses by the same factor, holding $a$ and $A$ fixed.  We find that orbital flips are preserved until $\delta$ is about $820$ times larger than the value shown in Eq.\ (\ref{eq3:params1}).    

\begin{figure}[t]
\begin{center}

\includegraphics[width=3.4in]{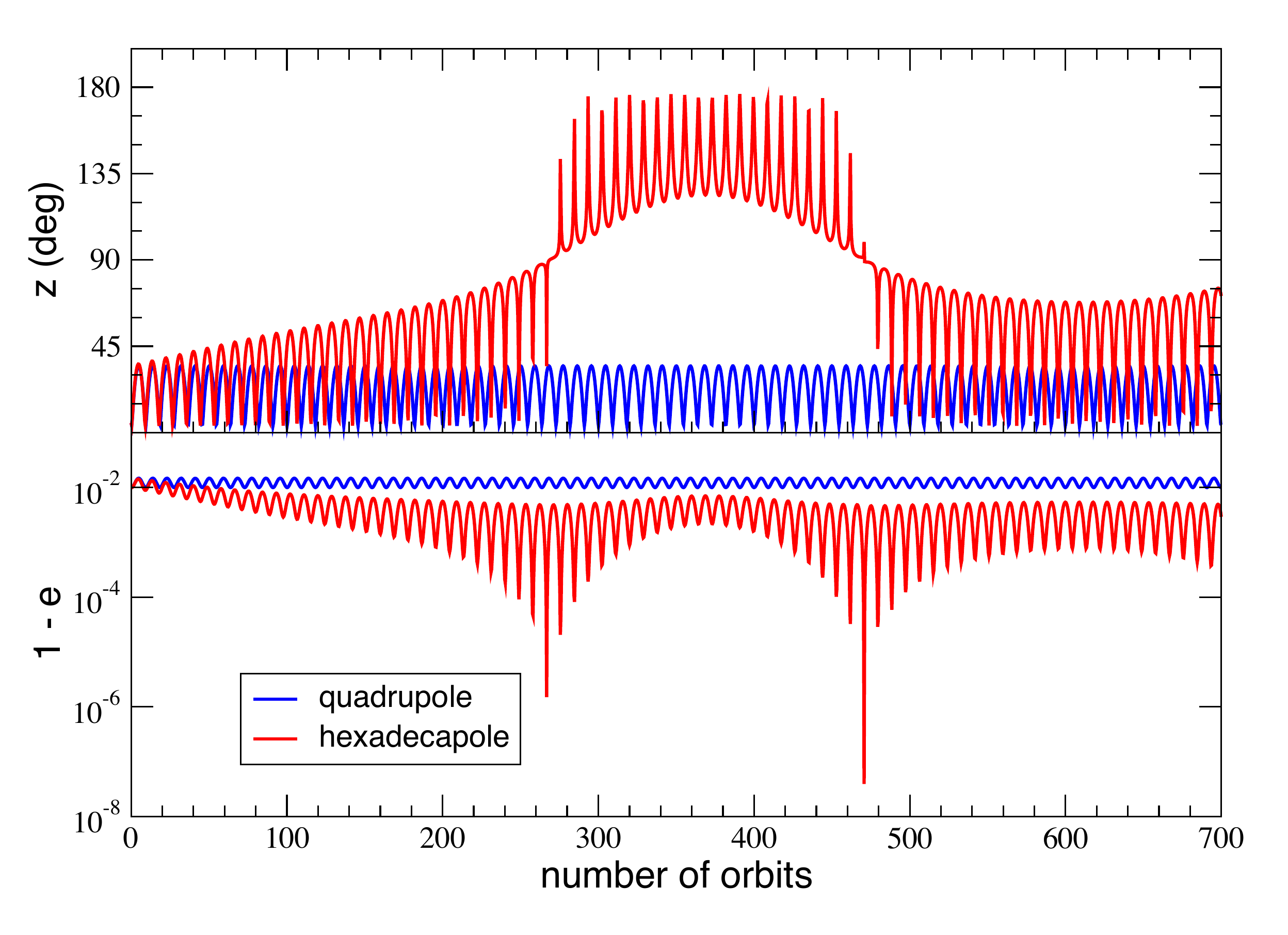}       

\caption{Evolution of $z$ and $e$ for an equal-mass inner binary in a nearly coplanar initial orbit. Blue: quadrupole order.  Red: hexadecapole order.  GR precessions are included.  (Color figures in online version.) \label{fig:EqualMasslowz} }
\end{center}
\end{figure}

Another example generates orbital flips from nearly coplanar orbits, an analogue of the case discussed in Sec.\ \ref{sec:coplanar}.   The results are shown in Fig.\ \ref{fig:EqualMasslowz}.  In this case the parameters are
\begin{align}
\alpha &= 17.857 \,, \quad \epsilon = 0.0875 \,, \quad   \delta = 3.95 \times 10^{-9} \,. 
\end{align}
A sample system is again two $1.4\, M_\odot$ neutron stars with $a = 7$ a.u., but now orbiting a $50 \, M_\odot$ star or black hole at $A=80$ a.u.  The initial conditions are 
\begin{equation}
e =  0.99 \,, \, E = 0.6 \,, \, z = 5^{\rm o} \,, \, \omega = 45^{\rm o} \,, 
\omega_3 = 0^{\rm o} \,.
\end{equation}
The quadrupole-order evolution, shown in blue, displays the standard Kozai-Lidov cycles, whereas the hexadecapole-order evolution shows orbital flips and excursions to extreme eccentricities, well beyond (in the sense of $\log (1-e)$) the initial relatively large initial value of $e=0.99$.   Here again, GR precessions play a negligible role, suppressing the flips only when the GR parameter $\delta$ is dialed up by a factor of about $80$.
 
\begin{figure}[t]
\begin{center}

\includegraphics[width=3.4in]{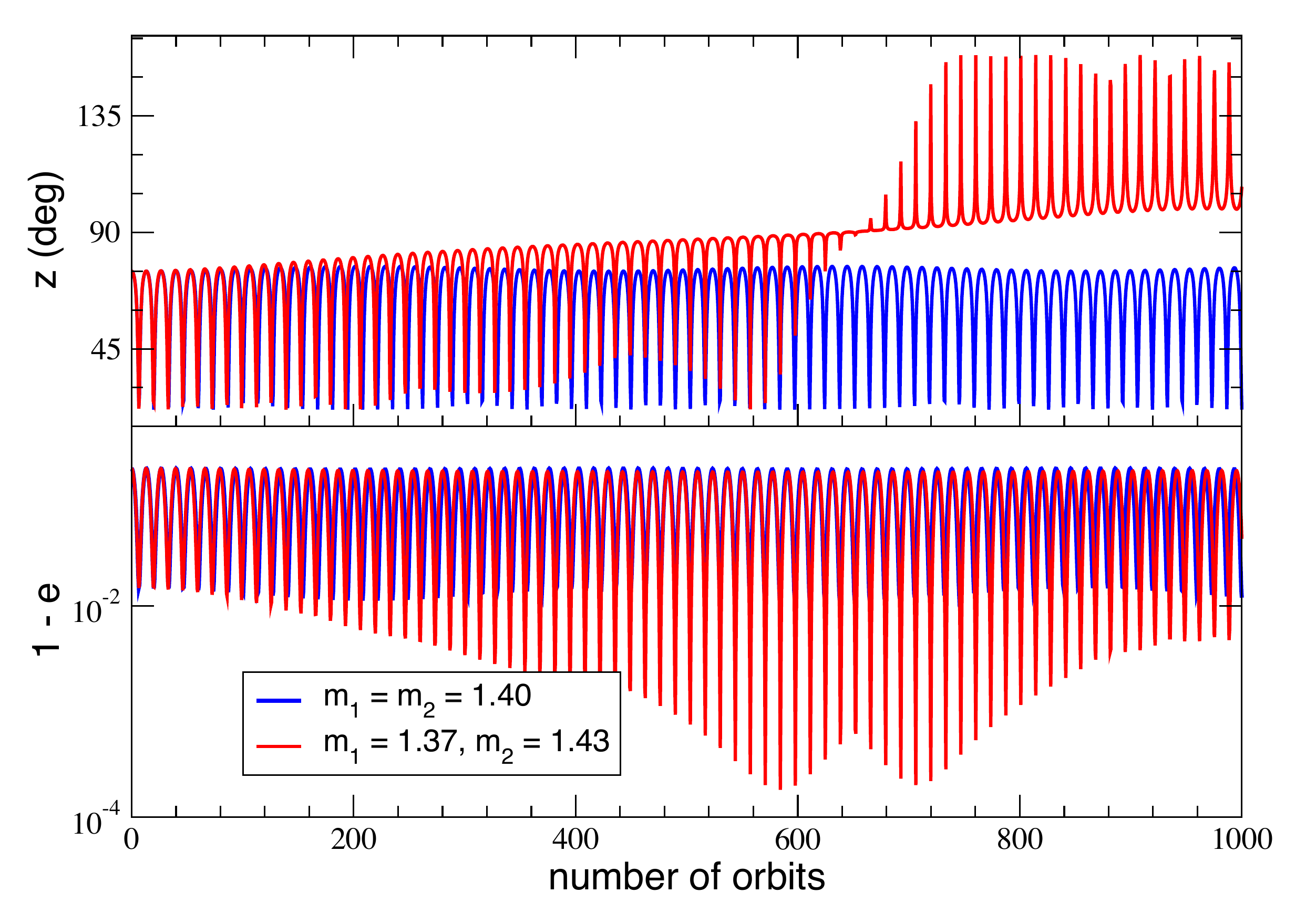}       

\caption{Evolution of $z$ and $e$ for equal and unequal-mass inner binaries  Blue: equal masses. Red: $m_1 = 1.33 \, M_\odot$, $m_2 = 1.47 \, M_\odot$.  GR precessions are included.  (Color figures in online version.) \label{fig:UnEqualMass} }
\end{center}
\end{figure}

As a final example, we display in Fig.\ \ref{fig:UnEqualMass} the effect of slightly unequal masses on the generation of orbital flips via octupole-order terms.   We again consider an inner binary of total mass $2.8 \, M_\odot$, with $a = 4$ a.u., orbiting a star or black hole of mass $50 \, M_\odot$ at $A = 50$ a.u.   The initial conditions are 
\begin{equation}
e =  0.8 \,, \, E = 0.6 \,, \, z = 75^{\rm o} \,, \, \omega = 0^{\rm o} \,, \,
\omega_3 = 0^{\rm o} \,.
\end{equation}
The equal-mass case shows no orbital flips in this case (blue curves in Fig.\ \ref{fig:UnEqualMass}), basically because $\epsilon = 0.08$ is smaller than in the previous cases, and the hexadecopole terms alone are not large enough to do the job.   As we change the two inner masses holding the total mass fixed, the octupole terms kick in, but are initially too small to generate flips, until we reach $m_1 = 1.33 \, M_\odot$, $m_2 = 1.47 \, M_\odot$, whereupon orbital flips and large eccentricities are generated (red curves).  

\section{Concluding remarks}
\label{sec:conclusions}

We have extended the study of Kozai-Lidov type hierarchical triple systems to hexadecapole order, or to order $(a/A)^5$, and examined a number of astrophysically interesting cases to elucidate the effect of the higher order terms on extreme behavior such as orbital flips and excursions to large eccentricity.   Given the complexity of the three-body problem, even in the hierarchical regime, it may come as no surprise that we find a complicated range of behaviors.  In most cases, the hexadecapole terms have only small quantitative effects on the long-term evolution of the system.  

In addition, in the astrophysical systems examined in Sec. \ref{sec:numerical}, the parameter $\epsilon$ ranged from $0.06$ to $0.24$; at the upper end of this range, the systems are not very hierarchical.  Given the inherently chaotic nature of the three-body problem, it pays to be cautious in ascribing a specific phenomenon (such as orbital flips) solely to the presence of a higher-order term, as opposed to a possible slight change in initial conditions.  

For equal-mass systems (and possibly for a range of nearly equal-mass systems), where the octupole terms vanish or are suppressed, we found a region of parameter space where orbital flips and excursions to large eccentricity could be generate by the hexadecapole terms.

We have derived and presented the equations in as clear a fashion as possible, to make it easy for other researchers to use them to explore the full parameter space of hierarchical triple behavior.  For example, all the examples discussed in this paper are characterized by $\beta \ll 1$, whereby the system's angular momentum resides primarily in the outer orbit.   The other limit, $\beta \gg 1$ may yield interesting behavior when hexadecapole terms are included (see \cite{2012ApJ...757...27A,2017arXiv170103795N} for a studies at octupole order).     Finally one should look at the interplay between these Newtonian $N$-body effects and GR effects beyond the basic pericenter precessions, including higher PN contributions, frame dragging effects, gravitational-radiation reaction damping, and effects arising from ``cross-terms'' between GR and quadrupole contributions  \cite{2014PhRvD..89d4043W}.

\acknowledgments

This work was supported in part by the National Science Foundation,
Grant Nos.\  PHY 16-00188.    We are particularly grateful to Smadar Naoz, Adrian Hamers, Gongjie Li, Jean Teyssandier, Fred Rasio and Todd Thompson for valuable comments on an earlier draft of this paper.   

\appendix

\section{A Delaunay/Osculating elements dictionary}
\label{app:dictionary}

Here we provide a dictionary that may be useful in translating between the language of osculating orbit elements used in this paper, and the language of Delaunay variables used in conventional treatments of many-body dynamics, and in particular in NFLRT \cite{2011Natur.473..187N,2013MNRAS.431.2155N}.

NFLRT used the subscript 2 to denote the orbit elements of the outer body, whereas we use the subscript 3; they use $k^2$ to denote the Newtonian constant $G$. 
There are six Delaunay ``coordinates'':  the two mean anomalies $\ell_1$ and $\ell_2$, which correspond roughly to our true anomalies $f$ and $F$, the longitudes of the ascending nodes $h_1$, and $h_2$, which correspond to $\Omega$ and $\Omega_3$ and the arguments of pericenter $g_1$ and $g_2$, which correspond to $\omega$ and $\omega_3$.
The ``conjugate momenta'' to those variables  are (Eqs.\ (3) -- (8) of \cite{2013MNRAS.431.2155N}):
\begin{align}
L_1 &= m \eta \sqrt{k^2 ma_1} \,, \quad L_2 = M \eta_3 \sqrt{k^2 M a_2} \,,
\nonumber\\
G_1 &= L_1 \sqrt{1-e_1^2}\,, \quad G_2 = L_2 \sqrt{1-e_2^2} \,,
\nonumber \\
H_1 &= G_1 \cos \iota_1 \,, \quad H_2 = G_2 \cos \iota_2 \,.
\end{align}
Since $G_1 = J_b = m\eta \sqrt{Gmp}$ and $G_2 = J_3 = M\eta_3 \sqrt{GMP}$,
it is straightforward to read off the correspondences $(a_1,\,a_2) \rightleftharpoons (a,\, A)$,
$(e_1,\, e_2)  \rightleftharpoons (e,\, E)$,
$(\iota_1,\, \iota_2)  \rightleftharpoons (\iota, \iota_3)$, with $z = \iota_{\rm tot} = \iota +\iota_3$.
Note that
\begin{equation}
\frac{G_1}{G_2} = \beta = \frac{\sin \iota_3}{\sin \iota} \,. 
\end{equation}

The parameters $C_2$ and $C_3$ of \cite{2013MNRAS.431.2155N}, Eqs. (21) and (B1),  are given by
\begin{align}
C_2 &= \frac{k^4}{16} \frac{m^7 m_3^7}{M^3 (m_1m_2)^3} \frac{L_1^4}{L_2^3 G_2^3} 
\nonumber \\
&= \frac{G}{16}
\eta \, \eta_3 \frac{M^2 a^2}{A^3 (1-E^2)^{3/2}} \,,
\nonumber \\
C_3 &= -\frac{15 k^4}{64} \frac{m^9 m_3^9 (m_1-m_2)}{M^4 (m_1m_2)^5} \frac{L_1^6}{L_2^3 G_2^5}
\nonumber \\
&
= \frac{15G}{64} \eta \, \eta_3 \Delta \frac{M^2 a^3}{A^4 (1-E^2)^{5/2}} \,.
\end{align}
The ratio
\begin{equation}
\frac{C_3}{C_2} = \frac{15}{4} \Delta \frac{a}{A} (1-E^2)^{-1} \,,
\end{equation}
consistent with Eq.\ (24) of \cite{2013MNRAS.431.2155N}.   
The amplitudes of the perturbing effects on the elements of each orbit are controlled in \cite{2013MNRAS.431.2155N} by the ratios
\begin{align}
\frac{C_2}{G_1} &= \frac{1}{16} \left (\frac{Gm}{a^3} \right )^{1/2}  \frac{\alpha \epsilon^3}{(1-e^2)^{1/2} (1-E^2)^{3/2}} \,,
\nonumber \\
\frac{C_2}{G_2} &= \frac{1}{16} \left (\frac{Gm}{a^3} \right )^{1/2}   \frac{\eta (1+\alpha)^{1/2} \epsilon^{7/2}}{(1-E^2)^{2}} \,,
\nonumber \\
\frac{C_3}{G_1}  &= \frac{15}{64} \left (\frac{Gm}{a^3} \right )^{1/2}  \frac{\alpha  \epsilon^4 \Delta}{(1-e^2)^{1/2} (1-E^2)^{5/2}} \,,
\nonumber \\
\frac{C_3}{G_2} &= \frac{15}{64} \left (\frac{Gm}{a^3} \right )^{1/2}   \frac{\eta (1+\alpha)^{1/2} \epsilon^{9/2} \Delta}{(1-E^2)^{3}} \,.
\end{align}
These amplitudes correspond to those displayed in 
Eqs.\ (\ref{eq2:quadrupole}) and (\ref{eq2:octupole}).  Finally, in making comparisons with \cite{2013MNRAS.431.2155N}, it is useful to note that 
\begin{align}
\frac{d\iota}{dt} &= \frac{\cos z \dot{G}_1 + \dot{G}_2 }{G_1 \sin z} \,,
\nonumber \\
\frac{d\iota_3}{dt} &= \frac{ \dot{G}_1 +\cos z \dot{G}_2 }{G_2 \sin z} \,.
\end{align}
With these translations, it can be shown that at quadrupole order, our Eqs.\ (\ref{eq2:quadrupole}) are identical to Eqs.\ (A26) - (A35), and that at octupole order, our Eqs.\ (\ref{eq2:octupole}) are identical to Eqs.\ (B6) - (B17) of \cite{2013MNRAS.431.2155N}.

\bibliography{refs}

\end{document}